\documentclass{article}

\usepackage{PRIMEarxiv}

\usepackage[utf8]{inputenc} 
\usepackage[T1]{fontenc}    
\usepackage{hyperref}       
\usepackage{url}            
\usepackage{booktabs}       
\usepackage{amsfonts}       
\usepackage{nicefrac}       
\usepackage{microtype}      
\usepackage{lipsum}
\usepackage{fancyhdr}       
\usepackage{graphicx}       
\graphicspath{{media/}}     

\title{Computer-Aided Cancer Diagnosis via Machine Learning and Deep Learning: A comparative review
}

\author{
  Solene Bechelli \\
  Biomedical Engineering \\
  University of North Dakota \\
  Grand Forks, ND\\
  \texttt{solene.bechelli@und.edu} \\
}

\begin{document}
\maketitle

\begin{abstract}
The past years have seen a considerable increase in cancer cases. However, a cancer diagnosis is often complex and depends on the types of images provided for analysis. It requires highly skilled practitioners but is often time-consuming and error-prone. If Machine Learning and deep learning algorithms have been widely used, a comprehensive review of the techniques used from the pre-processing steps to the final prediction is lacking. With this review, we aim to provide a comprehensive overview of the current steps required in building efficient and accurate machine learning algorithm for cancer prediction, detection and classification. To do so, we compile the results of cancer related study using AI over the past years. We include various cancers that encompass different types of images, and therefore different related techniques. We show that tremendous improvements have been made in the early detection of cancerous tumors and tissues. The techniques used are various and often problem-tailored and our findings is confirmed through the study of a large number of research. Moreover, we investigate the approaches best suited for different types of images such as histology, dermoscopic, MRI, etc. With this work, we summarize the main finding over the past years in cancer detection using deep learning techniques. We discuss the challenges of cancer research related to the large discrepancies in the images, and we provide some notable results in the field for lung, breast, and skin cancers.
\end{abstract}


\section{Introduction}
In the past decade, the number of computer-aided-diagnosis (CAD) studies via Machine Learning (ML) and deep learning (DL) algorithms has grown exponentially and has seen an incredible spike in their applications, especially in the biomedical field~\cite{brinker2019deep,assiri2020breast, manickavasagam2022cad, tandon2020putting, panourgias2006computational, munir2019cancer}. Their use in cancer detection is numerous and allows for rapid diagnosis of different cancer types. The world has seen an impressive increase in cancer cases. Not only the number has continuously grown; but with around 9 million cancer deaths in 2017 worldwide, and 2 million new cases and 600 thousand cancer deaths in 2021 in the United States alone ~\cite{siegel2021cancer}, the need for appropriate tools of detection and diagnosis is becoming more and more pressing both for accuracy and rapidity~\cite{ott2009importance}.
 
According to the American Cancer Society, the chances of survival over 5 years for an individual increase by 90\% when cancers are detected early. Similarly, screening for breast cancers has resulted in a lower death risk of 20-40\%~\cite{seely2018screening, berry2005effect}. However, lung cancers are detected in their later stages in 70 \% of the cases~\cite{svoboda2020deep}. In addition to decreasing chances of survival, late cancer detection leads to potential outbreaks of cancerous cells in other parts of the body leading to metastasis which needs to be prevented at all costs and provide a challenge for machine learning techniques.

Amongst all cancers, the tracheal, bronchus, and lung cancers are the most prevalent ones with a little under 2 million deaths, closely followed by colon, stomach, and liver cancers (digestive tract cancers) with around 800 thousand deaths for the year of 2017. Over the past decades, the number of deaths has decreased in the age group of 50-69 but increased in the age group of 70+ year old, which shows improvement in health care received for cancer treatments allowing patients to recover and live longer~\cite{siegel2021cancer}.

While treatments are becoming more effective for some cancer types, early detection is still the most successful approach to improve the chances of recovery.~\cite{miller2019cancer} Therefore, using machine learning and deep learning algorithms for diagnosis has gained traction in recent years. These algorithms allow for the rapid assessment of tumorous tissues with decreased human bias. They also lead to the earlier detection of cancers, {\it i.e.}, at stages for which tumors are sometimes difficult to analyze by the human eye.~\cite{houssein2021deep} Before discussing the different algorithms that have been developed to improve diagnosis, we briefly introduce the types of image that are often used in cancer detection as well as the three cancers chosen to conduct this work.

\subsection{Traditional cancer detection methods and image types\label{section:imageType}}

Cancer detection strongly depends on its type.~\cite{smith2002american, sharif2020comprehensive, la2020detection, warsavage2020quantifying} For instance, skin cancers can be diagnosed provisionally using dermoscopic images and non-invasive techniques. The use of non-invasive techniques is not always possible, and is often complex; therefore, diagnosis requires complementary methods such as biopsies. Using different techniques means that images with different specificities will need to be analyzed. In turn, it impacts the type of machine learning technique and the training process used to obtain a diagnosis.~\cite{fourcade2019deep} Hereafter, we give a quick overview of the different types of images primarily used in cancer detection and their related characteristics:

\begin{enumerate}
\item \textbf{Dermoscopic images}: These images are relevant to skin cancer detection. Dermoscopic images are direct images of a person's skin taken by a dermatologist for diagnosis purposes. The advantage of such an approach is that it is non-invasive. Nonetheless, images often have different characteristics due to various technical aspects such as camera angle, lighting, or contrast.\cite{sharif2020comprehensive} These differences make images complicated to analyze. In turn, images alone are often insufficient to lead to a correct diagnosis. However, such images are ideally suited to train Artificial Intelligence (AI) models. Performing a biopsy is required to confirm the diagnosis which, in turn, provides labeled databases.\cite{wen2021characteristics}
\item \textbf{X-Ray and CT scans}: Computerized Tomographic (CT) scans are often used in the early detection of lung and brain cancers (also found in some other cases).\cite{armato2011lung, rundo2021low, astaraki2021benign} They consist of a collection of X-ray images, taken from multiple angles, and have the advantage of providing a 3D rendering of tumors. CT scans allow for the rapid detection of abnormal growth. However, they can be expensive and complex to analyze. These grayscale images are often available from medical databases since CT scans are typical analyses for most cancer types. Since CT scans include many images, they readily lead to large training datasets which are necessary to train accurate models.\cite{armato2011lung} Moreover, these images have standardized characteristics, thus decreasing the number of pre-processing steps needed.
 \item \textbf{Histology}: Histology images are obtained from biopsies and provide direct insight into the cellular system of the patient. They allow for cancer detection at an early stage of the disease, before any physical manifestations ({\it i.e.} melanoma in skin cancers),\cite{arevalo2014histopathology} and are often used during breast cancer detection. The detection of cancerous tumors or tissues takes place on images with Hematoxylin and Eosin (H\&E) stains. The dye colors each image and allows to visualize cell nuclei and membranes. However, the intensity of the color depends on the amount of dye employed and is not standardized. These images are also very large, which complicates the analysis of large datasets. Histology images are of great significance as they allow for the determination of cancer stage and cancer treatment.\cite{roy2021computer}
\end{enumerate}

Deep learning techniques can also be applied to other types of images. These include Raman spectroscopy, mammography, ultrasound images, etc.\cite{houssein2021deep, gareau2020deep, jia2021breast, chen2019support} For instance, ultrasound images can be used for breast and thyroid cancer detection, and deep learning approaches have been developed for these applications.\cite{zhu2021generic} In this review, we focus on three specific types of cancer: lung, skin, and breast cancers since these cancers have been the most extensively studied in the biomedical field and provide a comprehensive access to the various techniques and image types.

\subsection{Types of cancers studied in the review}

\begin{figure}
    \centering
    \includegraphics{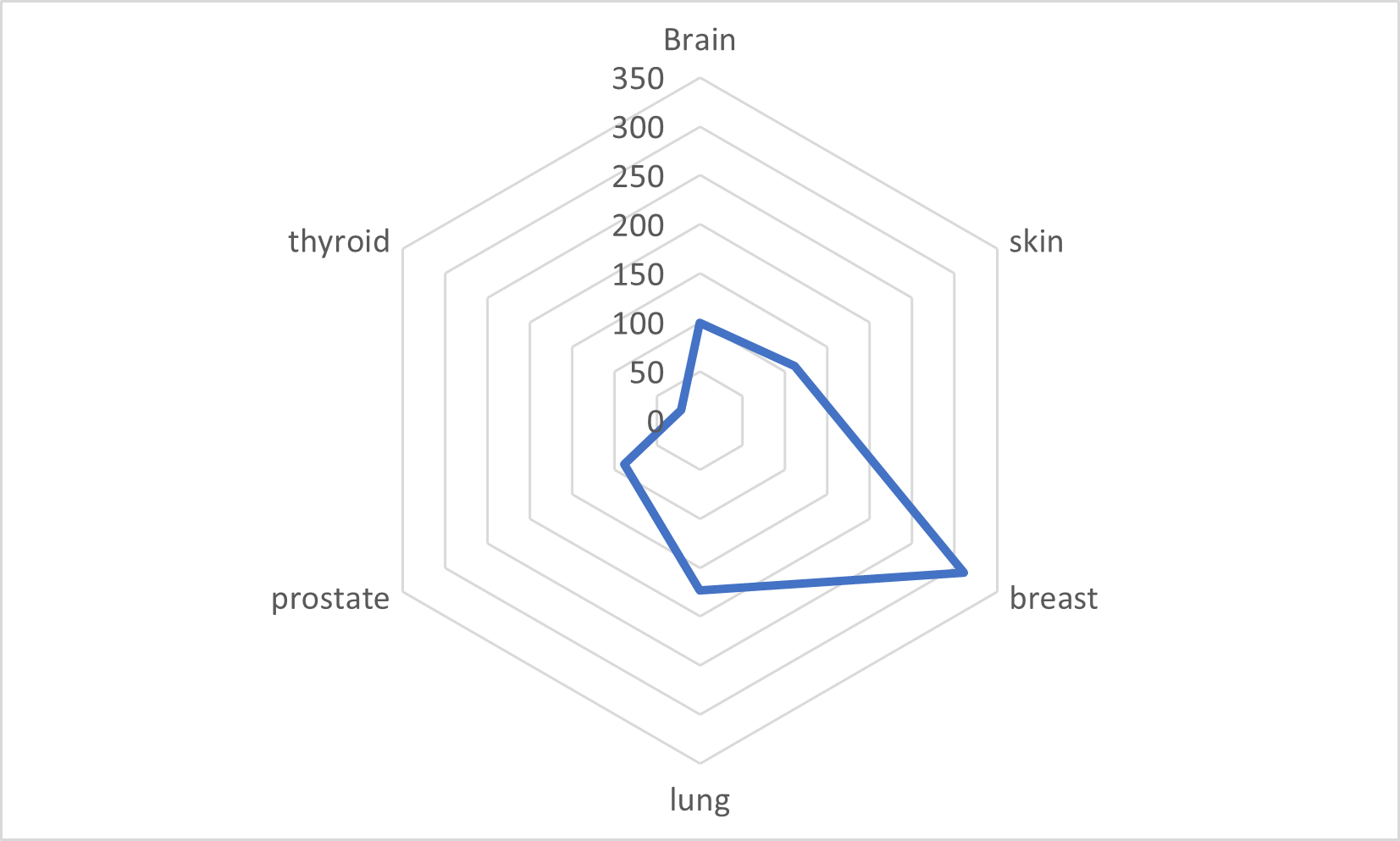}
    \caption{Number of papers in the PubMed database for different types of cancers over the 2017-2022 time frame. All papers involve the study of images with machine learning or deep learning tools.}
    \label{fig:cancer_studies}
\end{figure}

As discussed by Hanahan and Weinberg in their pioneering work~\cite{hanahan2000hallmarks}, the formation of tumors is a multistep process that results from genetic alterations. These genetic alterations lead to the gradual transformation of normal cells into malignant cells. The latter have several acquired capabilities, including their ability to be insensitive to antigrowth signals, mimic growth signals, evade apoptosis, self-replicate in a limitless manner, sustain angiogenesis and invade tissues and lead to metastasis. While some cancers, like leukemia, do not cause tumors to form, tumors are found commonly in many types of cancers, such as lung, skin, and breast cancers. In many cases, the presence of a tumor is the most visible symptom that initiates the diagnosis process~\cite{sharif2020comprehensive}. We focus here on tumorous cancers that provide adequate imaging access with different technical challenges. Moreover, since all nodules are not necessarily cancerous it increases the importance of diagnosis and appropriate tools to differentiate between the ones that are and the ones that are not cancerous.  As discussed in section \ref{section:imageType}, each type of cancer leads to different categories of images and, as a result, involves specific machine learning and deep learning approaches. In the case of lung cancer, many abnormalities may appear in the chest cavity and are often left untreated which, in turn, makes the diagnosis process complex~\cite{kim2020test}. For breast cancer, the challenge is to analyze both the tumor-like masses and also the onset of other processes, such as calcification, which are often early signs of the disease~\cite{blais2021deep}. Thus, we examine here studies performed primarily on breast, lung, and skin cancers. Moreover, figure \ref{fig:cancer_studies} shows an analysis of the studies published between 2017 and 2022. This plot shows that breast, lung, and skin cancers are the most studied and lend themselves to deep learning approaches.

The current literature is lacking a comprehensive overview of machine learning and deep techniques used to detect, classify or predict cancer tumors and tissues. Thus, in this review, we aim to offer a detailed analysis of the techniques used on various image and tumor types. This work aims to encompass various types of images and not limit itself to one specific image type. The contributions of this review are multiple and include: (i) a systematic approach to a large number of papers published in the recent years relating research in cancer detection using machine learning or deep learning, (ii) an overview of the pre-processing techniques commonly used for different image types, (iii) an analysis of the performance of the most common deep learning models and their classifier, and (iv) a comprehensive comparison of pre-trained models most often used in cancer diagnosis research. To do so, we start by presenting the review method used to select papers, including keywords, and inclusion and exclusion criteria. Then, we discuss the pre-processing and data augmentation techniques used to develop machine learning and deep learning models. Finally, we discuss the different feature extraction methods employed, from histogram analysis to convolutional layers, and detail the different classifiers so obtained and their applications. 

\section{Review method}
\subsection{Paper selection}

Papers for this study were selected using both the PubMed and BVS databases. A combination of the following keywords was used: “Machine learning, deep learning”, image classification”, “cancer detection” and “lung, breast and skin cancer” for both databases. After omitting books, we obtained a total of 582 scientific papers for the period between 2019 and 2021.

\subsection{Criteria for exclusion and inclusion}

We first removed all duplicates leading to a total of 363 remaining papers. We also discarded all articles for which full papers are not accessible, papers written in a different language from English, and book chapters. Then, some papers were removed on the basis of titles, which fell outside the scope of this review, leading to 234 papers. 

\section{Discussion}
Surveys and reviews have been published on a given type of cancer and the corresponding imaging technique. These include lung nodules from CT scans,\cite{tandon2020putting, mastouri2020deep, ather2020artificial, gu2021survey, mathew2020artificial,  wang2020research, liu2020evolving} dermoscopic images for skin cancers,\cite{aractingi2019computational, efimenko2020review, haggenmuller2021skin} and the application of AI for breast cancers diagnosis\cite{bitencourt2021ai, duggento2021deep, michael2021breast, ramadan2020methods, abdelhafiz2019deep, hosni2019reviewing, fujioka2020utility, ibrahim2020artificial, abdelrahman2021convolutional, gardezi2019breast, rashmi2022breast, zebari2021systematic}. Moreover, reviews have often focused on specific aspects, such as, {\it e.g.}, the comparison of transfer learning approaches~\cite{ayana2021transfer}, and a comprehensive review of the field is missing.\cite{munir2019cancer, debelee2020deep, ali2021state}

Previously published reviews have agreed on the increased use of deep learning in cancer detection methods.\cite{ather2020artificial, duggento2021deep} Therefore, we focus here on the applications of deep learning approaches in segmentation, detection, or classification/prediction tasks. We add that, in some cases, a combination of 2 (or more) deep learning methods is employed to improve performance.\cite{lee2021axillary} In the next section, we discuss all three aspects as they each have significant consequences on performance.

\begin{figure}
    \centering
    \includegraphics[width = 15cm]{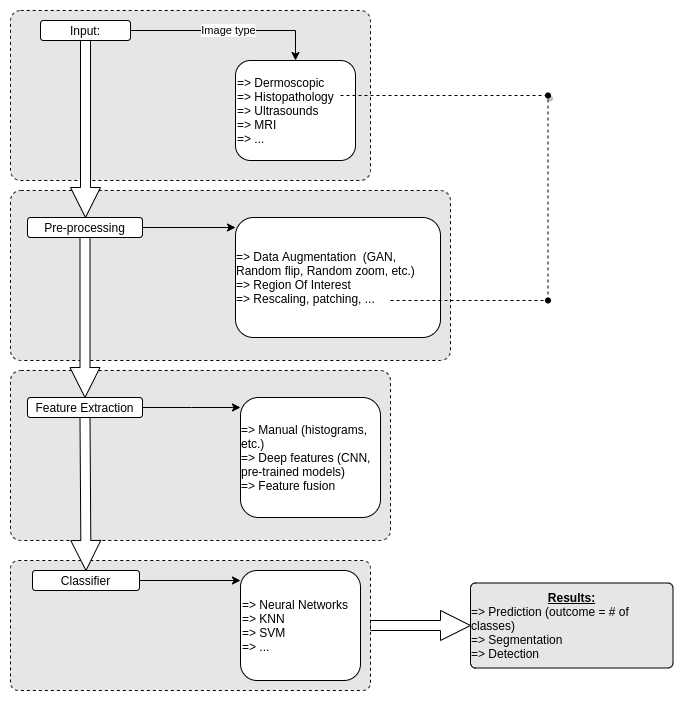}
    \caption{Workflow for deep learning and machine learning approaches to cancer detection.}
    \label{fig:workflow}
\end{figure}

The discussion follows the general workflow of deep learning algorithms shown in figure \ref{fig:workflow}. We first discuss the input type, and the pre-processing steps. We then move on to the feature extraction aspect of deep learning, before discussing the impact of classifiers on models' performance. We conclude with the analysis of some pre-trained models and additional general results.

\subsection{Pre-processing and image handling}
It is now widely recognized that the pre-processing of images has a significant impact on the results for all machine learning and deep learning approaches~\cite{ketabi2021computer}. However, the choice of a specific pre-processing technique depends on the image type, but also on the database~\cite{oyelade2021bioinspired,singla2019automated}. Therefore, the choice of the database is a significant step in the building of a machine learning/deep learning algorithm. An important analysis is done on the preparation of the database itself, which often comes in a non-practical format for machine learning usage. From handling labeling to pre-classification, this step allows for a properly built initial database. In addition, the database is often time separated into different categories to perform training, validation and testing of the model. A most common ratio being 80\% of the images to train the model and 20\% to test its performance.

Medical images are significantly different than regular images and, as such, require specific processing.\cite{shen2021interpretable} Indeed, low contrast images can lead to a difficult visualization of features.\cite{saffari2020fully} Pre-processing can include many different approaches: from handling large images,\cite{kimeswenger2021artificial, cao2019experimental,jaber2020deep, george2020breast} to using segmentation.\cite{chowdhary2020efficient, agnes2020appraisal, chaudhury2021breast, kassani2019comparative} In some cases, the images themselves need to be modified to be fully usable. An example of pre-processing operation is the removal of pectoral muscles in breast cancer studies~\cite{melekoodappattu2019hybridized}. Additional applications include contrast enhancement, color change, boundary extraction~\cite{saba2019region}, and noise reduction~\cite{wei2021optimized}. For instance, Boundary extraction of skin cancer was used by Saba et al. to increase the performance of their deep learning model on dermoscopic images~\cite{saba2019region}. Similarly, the use of noise reduction by Wei et al. improves the quality of the image fed to the network, thus enhancing the model's ability to differentiate between different cancer types~\cite{wei2021optimized}. Additionally, L$\hat{a}$ng {\it et al.} successfully used spatial and temporal subtraction to obtain a better contrast in breast cancer analysis~\cite{laang2021can}.

As discussed in section \ref{section:imageType}, histology images are often used for cancer diagnosis. The challenge here is to work with images of a gigabyte size, similar to those used for skin cancer analysis.\cite{kimeswenger2021artificial} A method that showed successful results when combined with CNN models is the patching technique.\cite{cao2019experimental,li2020multi, ciga2021overcoming, george2020computer} By reducing the image size, the work becomes less computationally intensive~\cite{jaber2020deep}. Indeed, patching of Whole Slide Imaging (WSI) can avoid deep segmentation, while still providing high accuracy, specificity, and sensitivity (96.91\%, 96.18\%, and 97.24\% respectively for the study of breast cancers~\cite{george2020breast}). Similar results were obtained on lung cancers.\cite{wang2019lung} Additionally, the large size of histological images can be decreased with segmentation steps.\cite{jojoa2021melanoma} However, new models have been developed recently to provide morphological features from WSI without patching,\cite{lu2021data} since even highly accurate models on patch-level can fail to identify the tumor bed correctly~\cite{ciga2021overcoming}. This technique has the advantage of removing some of the pre-processing steps, including finding the Region Of Interest (ROI)~\cite{cai2019breast, liu2020automatic, yu2020mammographic, johny2021dynamic}, In addition, it also seems to yield better results when combined with large structure models.\cite{ciga2021overcoming} 

In addition to the treatment of large images, medical databases often come with limitations, {\it e.g.}, a high imbalance between classes, or a lack of annotations. The pre-processing step often allows for the improvement of models' performances. For instance, skin cancer datasets are often imbalanced. To address this issue, mask-RCNN methods have been employed and have led to improvements in accuracy and other metrics~\cite{cao2021application}. Other studies on the example of breast cancer also report that pre-trained models can also solve this problem~\cite{singh2020imbalanced}.

Data augmentation is now widely used, as reflected by the literature. It allows for the extension of the original database. The application of the horizontal and vertical flipping of images improves the overall accuracy of a model~\cite{kassani2019comparative}. Similarly, data augmentation reduces the skewness of skin cancer datasets and their imbalance~\cite{kassani2019comparative}. Variations in dataset size and balance can strongly impact the accuracy of the model~\cite{khan2019construction}, and the benefits of using data augmentation have been highlighted in recent work~\cite{jojoa2021melanoma}. The lack of data in the medical field often results in incorrect diagnoses with a large number of false positives~\cite{wang2021false, duggento2019ad, yuan2021efficient}. To address this issue, novel data augmentation techniques, such as combinations of a graph convolutional network (GCN) and a convolution neural network (CNN), have been developed and have led to a significant improvement in the accuracy of the models~\cite{zhang2021improving, omigbodun2019effects}. Traditional Generative Adversarial Networks (GAN) algorithms can also be used to tailor data augmentation in skin cancer detection~\cite{roy2021computer, mutepfe2021generative}.
Similarly, additional data can be generated with a GAN-based radiomics model and TED-GAN algorithms~\cite{pang2021semi, rashid2019skin}.
The latter provides additional skin cancer images and improves the accuracy of the classifier from 66\% to 92.5\%~\cite{ahmad2021improving}. Moreover, data augmentation with ROI identification leads to significant improvements in the performance of CNNs, as shown by increases in accuracy and precision~\cite{ramadan2020using}. Similar conclusions hold for pre-trained models as well~\cite{sadad2021internet}. Another data augmentation technique, known as random cropping center (RCC),  also enhances the performance of deep learning models.\cite{wang2021boosted}

In addition, the scarcity of datasets with annotated images has slowed down progress. As a result, recent studies are focusing on segmentation and automatic annotation of X-ray images, which would be otherwise extremely time-consuming~\cite{wang2021automatic}. Alternative annotation techniques are currently under work to circumvent the problem. For instance, Lotter {\it et al.}'s method leverages localization-based information and out-performs~\cite{lotter2021robust}. Pre-processing also consists of retrieving annotations from the dataset, a time-consuming and costly task. To this end, the automatic labeling of histological images, powered by the use of confidence-boosting queries, has been developed~\cite{qi2018label}.

Along with offering labeling solutions, segmentation provides many possibilities for feature extraction~\cite{chaudhury2021breast, wang2019lung}, and is a crucial step in cancer detection~\cite{yusoff2020somatic, gu2019automatic, irfan2021dilated, kofler2021robust}. For instance, in breast cancer detection, one of the pre-processing steps consists of the identification, and sometimes removal, of pectoral muscles. Deep learning models have been used to this end in recent work~\cite{ma2019automated}. Systematic comparisons between ML approaches, with or without segmentation, have been performed. They show that the impact of segmentation correlates with the type of model used for classification~\cite{lee2021comparison}.
Higher success rates for diagnosis are obtained when segmentation and diagnosis tasks are two separate steps in the model~\cite{yu2021design, nasrullah2019automated, makrogiannis2021discriminative, li2021dual, yusoff2020somatic}, or when a succession of models, each dedicated to segmentation, detection, and classification process are employed~\cite{jojoa2021melanoma}. 
Separating segmentation and classification tasks into a combination of a GAN model for segmentation and a CNN for classification led to a 99.28\% specificity and a 98.75\% accuracy~\cite{saffari2020fully}. It allows for the identification of relevant areas for diagnosis and can involve different techniques, from masks to combinations with histograms~\cite{garcia2019segmentation, mahbod2020effects}. Results are often improved through contrast enhancement~\cite{chowdhary2020efficient,agnes2020appraisal, chaudhury2021breast, khan2019effective}. ROI extraction~\cite{yang2019comprehensive, melekoodappattu2019hybridized, kumar2021classification} increases the accuracy of CNN models as shown for skin cancer databases~\cite{zanddizari2021new}. Image quality can also be improved by reducing the background noise~\cite{sarath2020two}. In the case of WSI, the teams using the Otsu method for noise removal generally obtained better performance with accuracies up to 95 \%~\cite{liu2020evolving}. Similar results were obtained in studies on lung cancer studies~\cite{saba2019lung}. Optical-based methods also lead to accuracy above 97\% for breast cancer detection~\cite{kavitha2021deep} CNN models have also been used for segmentation purposes~\cite{piantadosi2020multi}. We add that the U-net model has been used on various types of medical images~\cite{yu2021design, blanc2020artificial, lu2021breast, jimenez2019deep, alom2019recurrent, kim2021computer, piantadosi2020multi}. Results show that its performance depends on the size of the nodules for lung cancer~\cite{yu2021design} and report 80\% accuracy on dermoscopic images for skin cancer diagnosis~\cite{kim2021computer}. For mammograms, segmentation is instrumental in locating the tumor~\cite{shen2020multicontext, agarwal2020deep}. Similarly, mask R-CNN is widely used to locate a nodule in lung cancers~\cite{nasrullah2019automated}, which is especially crucial when working on chest X-rays~\cite{li2020multi}.

\subsection{Feature extraction}

Feature extraction generally takes place after the pre-processing step. It is a crucial step in the development of high-performance models, which often rely on a large number of features. For instance, a dataset with 782 features was recently used in breast cancer classification~\cite{roy2021computer}. Assessing which features are the most relevant can be complex and has led to the design of feature-oriented techniques~\cite{pourasad2021presentation, sasikala2019particle, chang2020svm}. In this section, we focus on two types of feature extraction. First, we examine the case of low-level features and then turn to deep feature extraction. 

The type of low-level features extracted relies on a human choice. For instance, histogram extraction, via {\it e.g.}, HOG (histogram of gradients), is a common feature extracted for classification~\cite{kailasam2019novel}.
Examples of manually extracted features are fuzzy color histograms, RGB histograms, gray histograms, JPEG coefficient histograms, Haralick features, local binary partition histograms, and graph-based features. Low-level features can also include information on the shape, texture, border irregularities, and color of skin cancers~\cite{saba2019lung, chatterjee2019integration, li2019predicting,liu2020automatic}. To increase performance, algorithms such as SURF, Gabor filter (GFs), SIFT, ORB, and GLCM can also be used~\cite{melekoodappattu2019hybridized, roy2021computer}. Studies report an improvement in accuracy when radiomic features (shape of tumor/lesion, existence or lack of spicula, and texture of the sample) are used~\cite{sakai2020method, lei2019mammography}. Manually extracted features from local binary patterns and fed to an artificial neural network (ANN) can lead to a higher accuracy than some CNNs after deep-level feature extraction~\cite{alsaade2021developing}. However, manually extracted features are often not sufficient, particularly in skin cancer detection due to the similarity between lesions from one cancer to another~\cite{liu2020automatic}.

Deep learning models are often used to optimize the feature design tasks~\cite{manzo2020bucket}. Machine learning algorithms can provide low-accuracy predictions in skin cancer classification as many parameters need to be accounted for (light and angle, among others)~\cite{manzo2020bucket}. A CNN is often used in deep-level feature extraction. It is generally a great alternative to the manual extraction of features~\cite{ghazouani2021towards}, as shown by 3D convolutional neural networks for lung CT scan~\cite{ozdemir20193d}. They have been used to transform 2D color patches into classifiable 1D descriptive vectors~\cite{jaber2020deep}. Another advantage of feature extraction via deep learning is that it does not require any a priori knowledge and thus removes any potential bias~\cite{duggento2019ad}. Many other applications of CNN models as features extractor have been reported. This includes bio-inspired CNN models for breast cancer classification was bio-inspired which detect abnormalities in histological images~\cite{oyelade2021bioinspired}. Similarly, the BCHisto-Net model was developed to improve feature extraction on whole slide images in the case of breast cancer~\cite{rashmi2021bchisto}. However, deep-level features are generally associated with specific images and may not transfer to multiple classes easily~\cite{ameri2020deep}. Overall, feature extraction with CNNs is very efficient with results of 97\% accuracy reported for metastatic breast cancers~\cite{wang2021boosted}. A CNN also allows for the extraction of several features simultaneously. Khan {\it et al.} show that CNNs can lead to the identification of the three main characteristics of breast cancers (lump, masses, and micro-calcification)~\cite{heenaye2021multi} and report an 88\% accuracy for the classification between four types of cancers. Other studies have emphasized the significance of specific features, and the impact of the structure of the convolution-based model. Shu {\it et al.} show how the use of different types of pooling changes the accuracy of a CNN model for breast cancer detection~\cite{shu2020deep}. In addition, the use of a CNN for the extraction of relevant patches in mammography analysis, followed by the application of multiple-instance learning techniques, has led to an Area Under Curve (AUC) of 0.91~\cite{sarath2020two}. Transfer learning is also highly efficient for feature extraction and has led to very accurate results~\cite{chen2019computer, sharma2020conventional}.

In addition to low-level and deep feature extractions, prior work has developed approaches for the extraction of mid-level features~\cite{liu2020automatic}. 
While deep features (deep learning) and low-level handcrafted features (human learning) differ, one may obtain models of similar accuracy from these two types of features and learning~\cite{kimeswenger2021artificial}. Therefore, combining extraction methods at the two levels can provide better detection models~\cite{saba2019lung, li2019predicting}. For instance, recent work has shown that a combination of CNNs and handcrafted feature extraction techniques improves significantly the overall accuracy and addresses issues caused by small and imbalanced datasets~\cite{zhang2019investigation}. For breast cancer detection, such a combination leads to a 2\% increase in accuracy with respect to the sole application of deep feature extraction~\cite{daoud2020breast}. Similarly, multiple extraction features that leverage deep learning can be combined. For instance, a bi-linear convolutional neural network of VGG16 and VGG19 was proposed for the classification of lung cancers~\cite{mastouri2021bilinear} and breast cancers with a 94.2\% accuracy~\cite{daoud2020breast}. Fusions of CNNs with other pre-trained models, such as the ResNet50, for feature extraction, increased the AUC by 0.02 up to 0.94 for skin cancer detection~\cite{hagerty2019deep}. Feature extraction fusion has demonstrated a large improvement in models' performance for several types of cancer~\cite{ragab2021framework, zhuang2021breast, khan2019construction}. Dual pathways models have also been developed to study the impact of multiple features on the accuracy, and have led to improved results when compared to single pathway techniques~\cite{astaraki2021benign}. The combination of multiple feature types and the inclusion of parenchymal imaging features have also been found to result in increased accuracy~\cite{uthoff2019machine}.

Feature extraction may also consist of region extraction, for example, in the case of skin cancer~\cite{saba2019region}. The combination of patient metadata with a CNN model-based feature extraction greatly improves the accuracy of models for skin cancer detection (from 73.69\% to 92.34\% accuracy)~\cite{ningrum2021deep}. This requires having information on the patient ({\it e.g.}, age, gender, anatomical position, among others), and, in turn, predictions do not rely solely on the images fed to the model. Novel structures have recently been proposed~\cite{rashmi2021bchisto}. An example is a BCHisto-Net model for the analysis of breast images, which extracts both local and global features and out-performs state-of-the-art classifiers~\cite{rashmi2021bchisto}.

\subsection{Classifiers}

After feature extraction, various algorithms can be applied to perform classification tasks. This includes Support Vector Machine (SVM) and Neural Networks (NN) models, among others. The SVM classifier is typically used in the study of optical images for breast or lung cancers~\cite{yang2019comprehensive, yin2021prediction}. In recent years, machine learning methods were improved to compete with deep learning techniques. The so-called extreme learning machine method, with SVM and Naive Bayes (NB) classifiers, was devised to detect early calcification in breast cancer detection~\cite{melekoodappattu2019hybridized}. This technique provided over 99\% accuracy. SVM classifiers on handcrafted features led to 96\% in breast cancers~\cite{sasikala2019particle}, and a combination of CNN for deep feature extraction with an SVM for classification was applied to build a skin cancer prediction model~\cite{mahbod2019fusing}. A CART algorithm was developed for the analysis of Merckel cell carcinoma in skin cancer~\cite{cheraghlou2021machine}. The SVM classifier showed very good results in breast cancer detection on handcrafted radiomic features~\cite{sakai2020method}, on  MRI images~\cite{areiza2020novel}, but also on microwave images~\cite{rana2019machine}. A comparison of SVM, KNN, decision tree, and random forest (RF) classifiers for lung nodule classification shows that the decision tree is the least accurate, while SVM and KNN lead to an accuracy greater than 94\%~\cite{kailasam2019novel}. Similar results were obtained on other cancer types~\cite{chowdhary2020efficient}.
A study by Aswathy {\it et al.} used SVM for the classification of histology images of breast cancers. The results showed that SVM provided better accuracy (over 90\%) than NN and RF~\cite{aswathy2021svm}. A similar performance was also observed for lung nodule detection, with a 97.45\% sensitivity rate~\cite{khan2019effective}.
This method is able to provide a low number of false positives. Furthermore, an SVM classifier for the multi-classification of skin cancer yields an accuracy of 98.99\%, 97.54\%, and 99.65\% for data taken from different databases~\cite{chatterjee2019integration}. Finally, non-deep learning techniques, such as RF, have led to an accuracy of 83.51\% in studies on lung nodules~\cite{ding2021optical}. Other classifiers include attention classifiers~\cite{pesce2019learning, kimeswenger2021artificial}, in which attention networks are used to detect the position of lung lesions~\cite{pesce2019learning}, and adversarial training for skin cancer detection~\cite{zunair2020melanoma} and lung cancers~\cite{xie2019semi}.

The efficiency of classifiers generally depends on the type of cancer detection tasks and on the type of features extracted. For instance, in some cases, SVM classifiers can lead to a higher accuracy than softmax classifiers on lung nodules classification~\cite{mastouri2021bilinear}. Similarly, features extracted with direct convolutional neural networks and fed to a SVM classifier have been reported to be highly efficient for breast cancer classification~\cite{ragab2021framework,george2020breast}. Ensemble learning can also improve the performance of a classifier. Ensemble learning consists of assembling various machine learning classifiers and averaging their predictions~\cite{cao2021application, hosni2020mapping}.

Tailoring the characteristics of the classifier is often necessary to improve accuracy. Radiomic machine learning classifiers, implemented with an Elastic Net regularization for logistic regression, can double the rate of early detection of lung cancer patients with an AUC of 0.89~\cite{rundo2021low}. Similarly, classifiers can be optimized, {\it e.g.}, with the Whale optimization algorithm, and lead to an accuracy of 95\% for skin cancer classification~\cite{zhang2019optimization}. Loss function options are often explored to assess performance. For instance, Wang {\it et al.} provide an alternate loss function to that commonly used in softmax classifiers with a double-classifier~\cite{wang2021wdccnet}. Moreover, focal loss functions can improve accuracy (up to 97.2\%) for lung cancer classification, while retaining a similar sensitivity and specificity~\cite{tran2019improving}. The penalty loss function also showed great promise in reducing the number of false positives for lung cancer classification, leading to an accuracy of 91.60\%~\cite{xie2018knowledge}. These results emphasize the need for the appropriate tuning of a model. Johny {\it et al.} demonstrate the importance of fine-tuning CNN models, with dynamical learning rates resulting in better performance with a maximum accuracy of 91.84\%~\cite{johny2021dynamic}. 

\subsection{Transfer learning and pre-trained models}

Several studies leverage transfer learning via the use of pre-trained models. In general, pre-trained models are trained initially on the extensive ImageNet database~\cite{rv2021augmenting}. Pre-trained models are then applied to cancer detection and classification~\cite{saba2019region, cao2019experimental, george2020breast, ameri2020deep, zhang2020new, zhuang2021breast, suresh2019novel, cao2019experimental}. However, the ImageNet database does not focus on biomedical images. It, in turn, requires that the models be partially re-trained to be efficient for cancer detection. Several pre-trained models were recently developed for biomedical applications~\cite{gerard2018fissurenet, wang2021boosted, yang2021momminet}. These biomedically-oriented pre-trained models share a similar structure with popular pre-trained models, such as VGG16~\cite{montaha2021breastnet18}, and have been reported to yield an accuracy up to 98\%. To address the issue of generalization for models pre-trained on medical images, Gu {\it et al.} have proposed a two-step training. The model is first pre-trained on the ImageNet database and then fine-tuned on two different databases of dermoscopic images~\cite{gu2019progressive}. Moreover, Gu {\it et al.} showed that adversarial domain adaptation allows for better generalization. This approach also addresses issues arising from small and imbalanced databases by carrying out part of the training on a larger dataset. For instance, using Mask-DenseNet+ on an imbalanced dataset led to ~90\% accuracy and 0.95 AUC.\cite{cao2021application} 

Pre-trained models play a major role in all steps, from detection~\cite{traore2020evaluating} to segmentation. For instance, AlexNet and Resnet50 can be used for skin cancer detection~\cite{alsaade2021developing}, while ResNet50 can be used for the segmentation of lung nodules, leading to an accuracy of 87.3\%~\cite{yu2021design}. A study recently showed that region-based CNN (R-CNN) and RetinaNet modes provided the best performance for lung nodules detection~\cite{traore2020evaluating}. The accuracy was also shown to increase when combined with the patching method for breast cancer analysis~\cite{jaber2020deep}.

Even for a given cancer type, the performance of pre-trained models correlates with the image type. While models like EfficientNet-B3 are well suited to WSI-related diagnosis~\cite{ciga2021overcoming, abdolahi2020artificial}, others perform better on dermoscopic images. Similarly, studies indicate that VGG16 performs well for the binary classification of dermoscopic images~\cite{hasan2021comparative}, while ResNet50, Inception, and DenseNet provide better accuracy for multi-class prediction~\cite{singhal2020comparing}. An extensive comparison of pre-trained models established that they perform better than other deep learning and machine learning algorithms for classification tasks of lung cancer nodules using CT scans~\cite{yu2020reproducible}. In addition, pre-trained models were shown to perform very differently on mammogram analysis and tumor detection~\cite{yu2020mammographic}, which further demonstrates the impact of the model’s structure on feature extraction. For lung cancer detection, when combined with an SVM classifier, AlexNet has a higher specificity and sensitivity than GoogleNet~\cite{kumar2021classification}. In the case of WSI for breast cancer classification, AlexNet was found to outperform all previously reported results but required additional pre-processing steps~\cite{jimenez2019deep}.

Results also depend on the type of feature extracted before classification. Jain {\it et al.} compared multiple pre-trained models (VGG19, InceptionV3, InceptionResNetV2, ResNet50, Xception, and MobileNet) for skin cancer classification and found that Xception performed the best with a 90.48\% accuracy~\cite{jain2021deep}. Another comparison between pre-trained models (VGG19, ResNet50, Xception, DenseNet121, and EfficientNetB0) for skin cancer detection showed that EfficientNet was the most accurate with a 91.45\% accuracy~\cite{jaworek2021interpretability}. A combination of pre-trained models (ResNet-50, DensNet-201, and DarkNet-53) was reported to enhance the diagnosis of skin cancers, as this combination succeeded in extracting both low-level and deep features~\cite{attallah2021intelligent}. Ensemble networks are also often used with pre-trained models by combining different models, with a combination of ResNet152 and VGG19 leading to an accuracy of 90\%~\cite{tanaka2019computer}. Furthermore, MCDM (multi-criteria analysis method), which combines entropy and PROMETHEE (preference ranking organization method for enrichment of evaluations), has led to optimal performance for ResNet101 on skin cancer detection~\cite{alzahrani2021comprehensive}. The use of pre-trained models also comes with a limitation, as they often are CPU- or GPU-intensive to train and fine-tune~\cite{wu2020classification, abbas2021acral, shen2021interpretable, bioengineering9030097}. Furthermore, ensembles of pre-trained models, significantly improve the predictions' accuracy, as shown for breast cancer~\cite{alqudah2021sliding}. However, it does so at an even greater computing cost.

\subsection{Results}

We summarize in Table \ref{tab:results_CNN} results obtained with CNN models for skin cancer~\cite{brinker2019deep, saba2019region, jojoa2021melanoma, zhang2019optimization, song2020end, serte2020wavelet, foahom2021ensemble, nasiri2020depict, hasan2021comparative, bioengineering9030097}, lung cancer~\cite{yu2021design, aydin2021detection, tan2019expert, bonavita2020integration, ohno2020differentiation, nasrullah2019automated, zhang2019toward, manickavasagam2022cad} and breast cancer~\cite{lee2021axillary, jimenez2019deep, choudhury2021detecting, wahab2019transfer, sebai2020maskmitosis, abdolahi2020artificial, tanaka2019computer, zhou2019weakly, alom2019breast, zhang2020new, ciritsis2019automatic, borkowski2020fully}. Generally speaking, the CNN architecture is perfectly suited to obtain highly accurate results, with a 99\% accuracy reported in recent work~\cite{pourasad2021presentation}. CNN models have been instrumental in determining common patterns in histology images across cancer types~\cite{noorbakhsh2020deep}. They can be useful for both binary and multi-classification tasks, as discussed in recent work highlighting the importance of binary decisions~\cite{blais2021deep}, and the challenges of multi-class predictions~\cite{park2020deep}. Deep learning algorithms can determine subtypes of cancers, a diagnosis that is often more complex to make for an expert than the binary choice between malignant and benign~\cite{abbas2021acral}. Casting a multi-class classification into a binary classification can often alleviate this issue~\cite{foahom2021ensemble}. Results on lung and colon cancer classification reveal that models that perform well in binary decisions can exhibit a decreased performance in multi-class classification~\cite{masud2021machine}. 

\begin{figure}
    \centering
    \includegraphics{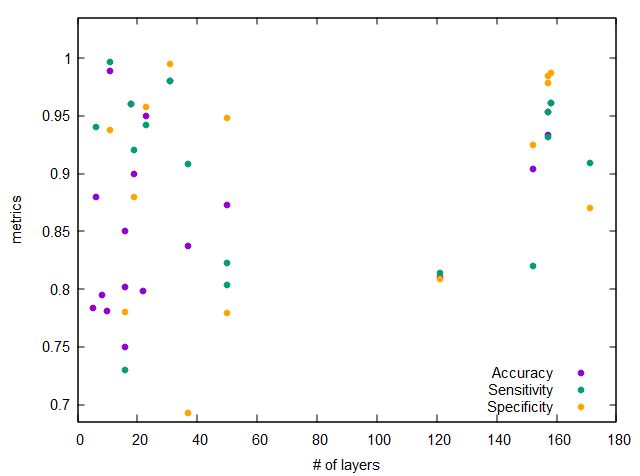}
    \caption{Metrics obtained in studies listed in Table~\ref{tab:results_CNN} against the number of layers of CNN and pre-trained models.}
    \label{fig:layer_metrics}
\end{figure}

Several studies have compared the performance of deep learning models with diagnoses made by specialists~\cite{lee2021axillary, nam2019development} or conventional image analyzers~\cite{sies2020past}. A comparison of the predictions made by CNN models with diagnoses from dermatologists shows that the specificity and sensitivity of deep learning models are superior~\cite{brinker2019deep}. The impact of several technical parameters on CNN models was assessed systematically in recent work. For instance, the assessment of the JPEG compression of WSI was evaluated and demonstrated the robustness of CNN models~\cite{zanjani2019impact}. The structure of the CNN network is also of great significance, as shown by Manikavasagam {\it et al.} who revealed that a five convolutional layer model leads to a 99\% accuracy ~\cite{manickavasagam2022cad}. Fig.~\ref{fig:layer_metrics} shows the impact of the number of layers on several metrics, including accuracy, specificity, and sensitivity. This plot also indicates that the model performance does not correlate with its size. Most models lead to an accuracy greater than 0.70. In addition, Fig.~\ref{fig:layer_metrics} shows that the deep learning models considered here are either very large, with more than a hundred layers, or fairly small, with fewer than 50 layers. Comparisons between CNN and pre-trained models have shown that small CNN tailored for a given application can perform better than the pre-trained models~\cite{lujan2021nanochest}. As discussed in the previous section, this has led to the development of biomedically-oriented pre-trained models trained on databases other than ImageNet.

\newgeometry{margin=0.5cm}

\begin{table}[]
    \centering
    \begin{tabular}{|c|c|c|c|c|c|c|c|}
    \hline
    \multicolumn{8}{|c|}{Skin Cancer}\\
    \hline
    Reference  &	Database    &	Accuracy    &	Precision   &	Recall=Sensitivity  &	AUC	    &   F-score   &   Specificity\\ 
    \hline
        \cite{brinker2019deep} & NA &           &               &  0.823            &          &               &   0.779  \\
        \cite{saba2019region}   &   PH2, ISBI   &   0.987, 0.951, 0.948 &   &   &   &   & \\
        \cite{jojoa2021melanoma} &  ISBI        &   0.904  &    &   0.820       &   &      &   0.925       \\
        \cite{zhang2019optimization}& DermIS, Dermquest  &   0.91    &    0.84     &   0.95    &       &          &       0.95    \\
        \cite{nasiri2020depict} &   ISIC    &   0.75    &   0.77    &   0.73    & 0.75  &   0.75   &   0.78    \\
    \hline
    \multicolumn{8}{|c|}{Lung Cancer}\\
    \hline
    Reference  &	Database    &	Accuracy    &	Precision   &	Recall=Sensitivity  &	AUC	    &   F-score   &   Specificity\\ 
    \hline
         \cite{manickavasagam2022cad} & LIDC/IDRI   & 0.9888  &       &   0.9962  &   0.928   &   &   0.9373    \\
         \cite{yu2021design}    & LIDC/IDRI  &   0.873   &   &   0.804   &   &   &   0.948   \\
         \cite{nasrullah2019automated}  &   LIDC/IDRI   & 0.8879  &   &   0.94    &   &   &   0.91  \\
         \cite{aydin2021detection}  &   NA  &  &    &   0.93    & &  0.87    &   0.82    \\ 
         \cite{tan2019expert}   &   LIDC/IDRI   &   0.88    &   & up to 0.94  &   0.943   &   &   \\   
         \cite{bonavita2020integration} &LIDC   &   &   0.83    &   0.83    &   &   0.83   &   \\
         \cite{ohno2020differentiation} & NA  & 0.90  &   & 0.92  &   0.94    &   &  0.88  \\
         \cite{zhang2019toward} & LUNA16  &   &   &   0.844   &   &   &   0.83    \\
    \hline
    \multicolumn{8}{|c|}{Breast Cancer}\\
    \hline
    Reference  &	Database    &	Accuracy    &	Precision   &	Recall=Sensitivity  &	AUC	    &   F-score   &   Specificity\\ 
    \hline
        \cite{lee2021axillary} &  NA  &   0.8105  & 0.7273  &   0.8136  &   0.8054  &  &   0.8085   \\
        \cite{jimenez2019deep}  & ICPR-2012, MITOS-ATYPIA-2014  &   0.95    &   &  0.9419 &   & 0.9435  & 0.9577  \\
        \cite{abdolahi2020artificial}   & kaggle  &   0.85    &  0.86 &   &   &   0.83    &   \\
        \cite{borkowski2020fully}   &   NA    &   0.96    &   0.92    &   0.96    &   &   0.94   &   \\
        \cite{ciritsis2019automatic}    & NA  &   &   &   0.8947 &   0.967  &    &   100 \\
        \cite{alom2019breast}       & BreakHis, BC 2015 $^{a}$ &   0.9751  & &   0.9452  &   0.9925  &   &   0.9829  \\
        \cite{zhou2019weakly}   &   NA  &   0.837   &   &   0.908   &   0.859  &   &   0.693   \\
        \cite{sebai2020maskmitosis} &   ICPR    &       &   0.921   &   0.811   &   &   0.863   &   \\
        \cite{choudhury2021detecting}   & NA  &   0.784   &   &   &   &   &   \\
        \cite{tanaka2019computer}   &   NA  &   &   &   0.909   &   0.951   &   &   0.870   \\
        \cite{zhang2020new}     & DDSM  &   0.9524  &   &   0.9611  &   0.9503  &   &   \\
    \hline
    \end{tabular}
    \caption{Summary of metrics reported for different CNN-based and pre-trained models for classification and detection of skin, lung and breast cancers. (a): Breast cancer challenge}
    \label{tab:results_CNN}
\end{table}

\restoregeometry

Several studies have focused on developing alternative CNN architectures. Recurrent Residual Convolutional Neural Network (IRRCNN) models exhibit higher accuracy than patching-based CNN models when applied to histological images~\cite{alom2019breast}, and approaches leveraging wavelet decomposition techniques led to a higher AUC, of the order of 94\% than conventional CNN models~\cite{masquelin2021wavelet}. The YOLO algorithm was also recently used in cancer detection and classification of mammograms~\cite{aly2021yolo, al2020evaluation}. Its combination with pre-trained models, such as ResNet50 for feature extraction, leads to high accuracy and an F1-score greater than 90\%~\cite{al2020evaluation}. While algorithms identify larger tumors accurately, masses that are close to one another are often challenging to diagnose~\cite{aly2021yolo}. Additional models that fuse the wider context of information using multiscale filters provide accurate results for the classification of breast cancers, with a sensitivity of 96\% and an AUC of 0.99, without increasing the computation time~\cite{agnes2020classification}. Similarly, radiomic-based models have drawn increased interest because of their ability to extract the most relevant features~\cite{chauvie2020artificial, binczyk2021radiomics}. Combinations of networks were recently developed, with an architecture composed of two independent artificial neural networks (ANNs), one feature constructor CNN, and one classification ANN. This led to the accurate identification of basal cell carcinoma in histological images~\cite{kimeswenger2021artificial}. Alternatively, deep residual neural networks, supplemented with transfer learning, allow for the differentiation of various lung cancer types from CT scans with up to 85\% accuracy.\cite{wang2020classification}. Deep residual neural network models are also used for cancer classification, especially when combined with SVM classifiers~\cite{wu2020classification,khan2019lungs}. 

Ensemble learning approaches are common techniques to improve the models' prediction~\cite{krajnc2021breast, manzo2020bucket, assiri2020breast,bioengineering9030097}. When applied to breast cancer classification, ensembles of pre-trained models achieve up to 90\% accuracy~\cite{barsha2021automated}. Similarly, when combined with directed acyclic graphs, ensemble learning yields a 76.6\% accuracy for the multi-class classification of dermoscopic images~\cite{foahom2021ensemble}.

Several studies have also examined the potential of deep learning to obtain parallel information that leads to cancer diagnosis. For instance, the cancer stage correlates with the nodule size. Ohno {\it et al.} use CNNs to determine the change in nodule size in lung cancers~\cite{ohno2020differentiation}. Deep learning techniques can calculate the number of nuclei and quantify the spread of cancerous cells~\cite{krijgsman2020quantitative}. However, it is important to note that the accuracy reported in classification studies that require prior knowledge of the nodules' position will likely decrease if both detection and classification tasks were carried out~\cite{bonavita2020integration}. It illustrates the limitation of traditional CNN models and accounts for the recent development of segmentation and detection-based models. 

CNN models also require large datasets with labeled images. Such datasets are often hard to access but are necessary to perform supervised learning. Thus, partly supervised or fully unsupervised techniques have recently been designed for cancer classification~\cite{xie2019semi, johny2021dynamic}. The former focuses on semi-supervised networks and leads to a 92.53\% accuracy. However, supervised learning has recently been reported to be more selective than unsupervised learning for lung nodule classification~\cite{astaraki2021benign}.

The choice of a specific dataset is often crucial and leads to significant accuracy differences (94.8\% to 98.4\% on skin cancers)~\cite{saba2019region}. Furthermore, methods were recently proposed to reduce the number of false positives~\cite{loizidou2021digital} by removing non-relevant images with R-CNN models~\cite{liang2021reducing} and through the use of cascade models, with a detection step followed by a diagnosis step~\cite{ozdemir20193d}. A drawback of the first approach is that it reduces the number of images in the database. While the second approach does not reduce the database size, it results in an increase in computing time. This is because two or more models are required, and this can become a significant issue for applications involving large databases~\cite{saba2019automated}. Alternatively, combining the model’s prediction with physiological and clinical information can also lead to a reduction in the number of false positives~\cite{saba2019automated}. Dual-path techniques for CNN (DualCoreNet) have shown promising results by combining segmentation and classification tasks~\cite{li2021dual}. Given the general scarcity of databases with labeled data, unsupervised learning methods are emerging, but have yet to become popular in the biomedical field~\cite{santilli2021domain, sebai2020maskmitosis, kofler2021robust}. 

Metrics, such as sensitivity and specificity rather than accuracy, allow for a direct comparison with diagnoses made by physicians~\cite{hosking2019hyperspectral}. As a result, studies are increasingly focusing on the improvement of specificity and sensitivity rather than accuracy~\cite{wu2019machine, saba2019automated, zhang2019toward, zhang2021improving}. We add that other metrics such as the F2- and F3-score are also valuable to assess the model's performance~\cite{assiri2020breast}.

\section{Conclusion}
In this review, we discuss the tremendous, and ongoing, progress in applications of machine learning and deep learning for cancer detection and diagnosis. We focus primarily on the examples of medical images for lung, skin, and breast cancer detection. Here we examine the various steps involved in the building of a cancer prediction model, from the pre-processing of the data, the extraction of features, and the classification tasks. The application of machine learning and deep learning methods in the biomedical field is extremely bright. Comparisons between diagnoses made by pathologists and models' predictions reveal that such models can lead to a higher diagnostic accuracy than specialists~\cite{marchetti2020computer, maron2020artificial, wu2019deep, choi2021evaluation}. However, this review also identifies several outstanding challenges. Model training is often limited to a single database and, as such, relies on a rather small number of images that originate from the same source and thus lack diversity~\cite{saffari2020fully, montaha2021breastnet18}. This means that generalization to unseen data remains a significant challenge~\cite{wang2020inconsistent}. Another widespread issue is the lack of public availability of datasets developed in-house~\cite{yang2021momminet}. Furthermore, ongoing efforts focus on improving the efficiency of segmentation and detection tasks~\cite{tennakoon2019classification, kofler2021robust}.

The applications of machine learning and deep learning algorithms extend well beyond the development of cancer prediction models in "simple" cases, {\it i.e.}, when no other pathology can account for the symptoms. In recent work, DL models can classify and diagnose cancers, COVID-19, and pneumonia from chest X-rays and CT images~\cite{ibrahim2021deep}. Deep learning is now applied to uncover connections between histological images and multi-omic features~\cite{ektefaie2021integrative}. Further insight can also be provided by applying deep learning to other types of data, including clinical studies~\cite{kehl2019assessment}, cancer-related diseases~\cite{chowdhary2020efficient}, actin fibers detection,\cite{oei2019convolutional}, volumetric data~\cite{tennakoon2019classification}, as well as genetic information~\cite{wang2021deep, wang2021prediction, anand2020deep, al2021nanogenomics}. Deep learning has also shown promising results in treatment prediction, risk prediction, and lab analysis~\cite{han2020augmented, ho2021detecting,laang2021can}. Finally, the handling of personal data and medical history will require secure protocols. Protocols and software are currently under development for applications in a hospital setting~\cite{kumar2021integration}, but also in the case of smartphone apps for skin cancer detection~\cite{udrea2020accuracy, jiang2020recognizing}.\\

\section*{Acknowledgments}

Partial funding for this research was supported by the National Science Foundation under NSF EPSCoR Track-1 Cooperative Agreement OIA \# 1946202.

\bibliographystyle{unsrt}  
\bibliography{references}

\begin{thebibliography}{100}

\bibitem{brinker2019deep}
Titus~J Brinker, Achim Hekler, Alexander~H Enk, Carola Berking, Sebastian
  Haferkamp, Axel Hauschild, Michael Weichenthal, Joachim Klode, Dirk
  Schadendorf, Tim Holland-Letz, et~al.
\newblock Deep neural networks are superior to dermatologists in melanoma image
  classification.
\newblock {\em Eur. J. Cancer}, 119:11--17, 2019.

\bibitem{assiri2020breast}
Adel~S Assiri, Saima Nazir, and Sergio~A Velastin.
\newblock Breast tumor classification using an ensemble machine learning
  method.
\newblock {\em J. Imaging}, 6(6):39, 2020.

\bibitem{manickavasagam2022cad}
R~Manickavasagam, S~Selvan, and Mary Selvan.
\newblock Cad system for lung nodule detection using deep learning with cnn.
\newblock {\em Med. Biol. Eng. Comput.}, 60(1):221--228, 2022.

\bibitem{tandon2020putting}
Yasmeen~K Tandon, Brian~J Bartholmai, and Chi~Wan Koo.
\newblock Putting artificial intelligence (ai) on the spot: machine learning
  evaluation of pulmonary nodules.
\newblock {\em J. Thorac. Dis.}, 12(11):6954, 2020.

\bibitem{panourgias2006computational}
Evangelia Panourgias, Athanasios Tsakonas, Georgios Dounias, and Georgia
  Panagi.
\newblock Computational intelligence for the detection and classification of
  malignant lesions in screening mammography.
\newblock {\em Oncol. Rep.}, 15(4):1037--1041, 2006.

\bibitem{munir2019cancer}
Khushboo Munir, Hassan Elahi, Afsheen Ayub, Fabrizio Frezza, and Antonello
  Rizzi.
\newblock Cancer diagnosis using deep learning: a bibliographic review.
\newblock {\em Cancers}, 11(9):1235, 2019.

\bibitem{siegel2021cancer}
Rebecca~L Siegel, Kimberly~D Miller, Hannah~E Fuchs, and Ahmedin Jemal.
\newblock Cancer statistics, 2021.
\newblock {\em CA Cancer J. Clin.}, 71(1):7--33, 2021.

\bibitem{ott2009importance}
JJ~Ott, A~Ullrich, and AB~Miller.
\newblock The importance of early symptom recognition in the context of early
  detection and cancer survival.
\newblock {\em Eur. J. Cancer}, 45(16):2743--2748, 2009.

\bibitem{seely2018screening}
JM~Seely and T~Alhassan.
\newblock Screening for breast cancer in 2018—what should we be doing today?
\newblock {\em Curr. Oncol.}, 25(s1):115--124, 2018.

\bibitem{berry2005effect}
Donald~A Berry, Kathleen~A Cronin, Sylvia~K Plevritis, Dennis~G Fryback, Lauren
  Clarke, Marvin Zelen, Jeanne~S Mandelblatt, Andrei~Y Yakovlev, J~Dik~F
  Habbema, and Eric~J Feuer.
\newblock Effect of screening and adjuvant therapy on mortality from breast
  cancer.
\newblock {\em New Engl. J. Med.}, 353(17):1784--1792, 2005.

\bibitem{svoboda2020deep}
Elizabeth Svoboda.
\newblock Deep learning delivers early detection.
\newblock {\em Nature}, 587(7834):S20--S22, 2020.

\bibitem{miller2019cancer}
Kimberly~D Miller, Leticia Nogueira, Angela~B Mariotto, Julia~H Rowland,
  K~Robin Yabroff, Catherine~M Alfano, Ahmedin Jemal, Joan~L Kramer, and
  Rebecca~L Siegel.
\newblock Cancer treatment and survivorship statistics, 2019.
\newblock {\em CA Cancer J. Clin.}, 69(5):363--385, 2019.

\bibitem{houssein2021deep}
Essam~H Houssein, Marwa~M Emam, Abdelmgeid~A Ali, and Ponnuthurai~Nagaratnam
  Suganthan.
\newblock Deep and machine learning techniques for medical imaging-based breast
  cancer: A comprehensive review.
\newblock {\em Expert. Sys. Appl.}, 167:114161, 2021.

\bibitem{smith2002american}
Robert~A Smith, Vilma Cokkinides, Andrew~C von Eschenbach, Bernard Levin,
  Carmel Cohen, Carolyn~D Runowicz, Stephen Sener, Debbie Saslow, and Harmon~J
  Eyre.
\newblock American cancer society guidelines for the early detection of cancer.
\newblock {\em CA Cancer J. Clin.}, 52(1):8--22, 2002.

\bibitem{sharif2020comprehensive}
Muhammad~Irfan Sharif, Jian~Ping Li, Javeria Naz, and Iqra Rashid.
\newblock A comprehensive review on multi-organs tumor detection based on
  machine learning.
\newblock {\em Pattern Recogn. Lett.}, 131:30--37, 2020.

\bibitem{la2020detection}
David La~Barbera, Ant{\'o}nio Pol{\'o}nia, Kevin Roitero, Eduardo Conde-Sousa,
  and Vincenzo Della~Mea.
\newblock Detection of her2 from haematoxylin-eosin slides through a cascade of
  deep learning classifiers via multi-instance learning.
\newblock {\em J. Imaging}, 6(9):82, 2020.

\bibitem{warsavage2020quantifying}
Theodore Warsavage~Jr, Fuyong Xing, Anna~E Bar{\'o}n, William~J Feser, Erin
  Hirsch, York~E Miller, Stephen Malkoski, Holly~J Wolf, David~O Wilson, and
  Debashis Ghosh.
\newblock Quantifying the incremental value of deep learning: Application to
  lung nodule detection.
\newblock {\em PLoS one}, 15(4):e0231468, 2020.

\bibitem{fourcade2019deep}
A~Fourcade and RH~Khonsari.
\newblock Deep learning in medical image analysis: A third eye for doctors.
\newblock {\em J. Stomatol. Oral Maxillofac. Surg.}, 120(4):279--288, 2019.

\bibitem{wen2021characteristics}
David Wen, Saad~M Khan, Antonio~Ji Xu, Hussein Ibrahim, Luke Smith, Jose
  Caballero, Luis Zepeda, Carlos de~Blas~Perez, Alastair~K Denniston, Xiaoxuan
  Liu, et~al.
\newblock Characteristics of publicly available skin cancer image datasets: a
  systematic review.
\newblock {\em Lancet Digit. Health}, 2021.

\bibitem{armato2011lung}
Samuel~G Armato~III, Geoffrey McLennan, Luc Bidaut, Michael~F McNitt-Gray,
  Charles~R Meyer, Anthony~P Reeves, Binsheng Zhao, Denise~R Aberle, Claudia~I
  Henschke, Eric~A Hoffman, et~al.
\newblock The lung image database consortium (lidc) and image database resource
  initiative (idri): a completed reference database of lung nodules on ct
  scans.
\newblock {\em Med. Phys.}, 38(2):915--931, 2011.

\bibitem{rundo2021low}
Leonardo Rundo, Roberta~Eufrasia Ledda, Christian di~Noia, Evis Sala, Giancarlo
  Mauri, Gianluca Milanese, Nicola Sverzellati, Giovanni Apolone, Maria~Carla
  Gilardi, Maria~Cristina Messa, et~al.
\newblock A low-dose ct-based radiomic model to improve characterization and
  screening recall intervals of indeterminate prevalent pulmonary nodules.
\newblock {\em Diagnostics}, 11(9):1610, 2021.

\bibitem{astaraki2021benign}
Mehdi Astaraki, Yousuf Zakko, Iuliana~Toma Dasu, {\"O}rjan Smedby, and
  Chunliang Wang.
\newblock Benign-malignant pulmonary nodule classification in low-dose ct with
  convolutional features.
\newblock {\em Phys. Med.}, 83:146--153, 2021.

\bibitem{arevalo2014histopathology}
John Arevalo, Angel Cruz-Roa, and FABIO~A GONZ{\'A}LEZ~O.
\newblock Histopathology image representation for automatic analysis: A
  state-of-the-art review.
\newblock {\em Rev. Med.}, 22(2):79--91, 2014.

\bibitem{roy2021computer}
Soumya~Deep Roy, Soham Das, Devroop Kar, Friedhelm Schwenker, and Ram Sarkar.
\newblock Computer aided breast cancer detection using ensembling of texture
  and statistical image features.
\newblock {\em Sensors}, 21(11):3628, 2021.

\bibitem{gareau2020deep}
Daniel~S Gareau, James Browning, Joel~Correa Da~Rosa, Mayte Suarez-Farinas,
  Samantha Lish, Amanda~M Zong, Benjamin Firester, Charles Vrattos, Yael
  Renert-Yuval, Mauricio Gamboa, et~al.
\newblock Deep learning-level melanoma detection by interpretable machine
  learning and imaging biomarker cues.
\newblock {\em J. Biomed. Opt.}, 25(11):112906, 2020.

\bibitem{jia2021breast}
Dongfang Jia, Cheng Chen, Chen Chen, Fangfang Chen, Ningrui Zhang, Ziwei Yan,
  and Xiaoyi Lv.
\newblock Breast cancer case identification based on deep learning and
  bioinformatics analysis.
\newblock {\em Front. Genet.}, 12, 2021.

\bibitem{chen2019support}
Bingling Chen, Yuan Lu, Wenhui Pan, Jia Xiong, Zhigang Yang, Wei Yan, Liwei
  Liu, and Junle Qu.
\newblock Support vector machine classification of nonmelanoma skin lesions
  based on fluorescence lifetime imaging microscopy.
\newblock {\em Anal. Chem.}, 91(16):10640--10647, 2019.

\bibitem{zhu2021generic}
Yi-Cheng Zhu, Alaa AlZoubi, Sabah Jassim, Quan Jiang, Yuan Zhang, Yong-Bing
  Wang, Xian-De Ye, and DU~Hongbo.
\newblock A generic deep learning framework to classify thyroid and breast
  lesions in ultrasound images.
\newblock {\em Ultrasonics}, 110:106300, 2021.

\bibitem{hanahan2000hallmarks}
Douglas Hanahan and Robert~A Weinberg.
\newblock The hallmarks of cancer.
\newblock {\em cell}, 100(1):57--70, 2000.

\bibitem{kim2020test}
Hyungjin Kim, Chang~Min Park, and Jin~Mo Goo.
\newblock Test-retest reproducibility of a deep learning--based automatic
  detection algorithm for the chest radiograph.
\newblock {\em Eur. Radiol.}, 30(4):2346--2355, 2020.

\bibitem{blais2021deep}
Marc-Andr{\'e} Blais and Moulay~A Akhloufi.
\newblock Deep learning and binary relevance classification of multiple
  diseases using chest x-ray images.
\newblock In {\em 2021 43rd Annual International Conference of the IEEE
  Engineering in Medicine \& Biology Society (EMBC)}, pages 2794--2797. IEEE,
  2021.

\bibitem{mastouri2020deep}
Rekka Mastouri, Nawres Khlifa, Henda Neji, and Saoussen Hantous-Zannad.
\newblock Deep learning-based cad schemes for the detection and classification
  of lung nodules from ct images: A survey.
\newblock {\em J. X-ray Sci. Technol.}, 28(4):591--617, 2020.

\bibitem{ather2020artificial}
S~Ather, T~Kadir, and F~Gleeson.
\newblock Artificial intelligence and radiomics in pulmonary nodule management:
  current status and future applications.
\newblock {\em Clin. Radiol.}, 75(1):13--19, 2020.

\bibitem{gu2021survey}
Yu~Gu, Jingqian Chi, Jiaqi Liu, Lidong Yang, Baohua Zhang, Dahua Yu, Ying Zhao,
  and Xiaoqi Lu.
\newblock A survey of computer-aided diagnosis of lung nodules from ct scans
  using deep learning.
\newblock {\em Comput. Biol. Med.}, 137:104806, 2021.

\bibitem{mathew2020artificial}
Christopher~Joy Mathew, Ashwini~Maria David, and Chris Mariya~Joy Mathew.
\newblock Artificial intelligence and its future potential in lung cancer
  screening.
\newblock {\em EXCLI J.}, 19:1552, 2020.

\bibitem{wang2020research}
Yu~Wang, Bo~Wu, Nan Zhang, Jiabao Liu, Fei Ren, and Liqin Zhao.
\newblock Research progress of computer aided diagnosis system for pulmonary
  nodules in ct images.
\newblock {\em J. X-ray Sci. Technol.}, 28(1):1--16, 2020.

\bibitem{liu2020evolving}
Bo~Liu, Wenhao Chi, Xinran Li, Peng Li, Wenhua Liang, Haiping Liu, Wei Wang,
  and Jianxing He.
\newblock Evolving the pulmonary nodules diagnosis from classical approaches to
  deep learning-aided decision support: three decades’ development course and
  future prospect.
\newblock {\em J. Cancer Res. Clin. Oncol.}, 146(1):153--185, 2020.

\bibitem{aractingi2019computational}
Selim Aractingi and Giovanni Pellacani.
\newblock Computational neural network in melanocytic lesions diagnosis:
  artificial intelligence to improve diagnosis in dermatology?
\newblock {\em Eur. J. Dermatol.}, 29(1):4--7, 2019.

\bibitem{efimenko2020review}
Mila Efimenko, Alexander Ignatev, and Konstantin Koshechkin.
\newblock Review of medical image recognition technologies to detect melanomas
  using neural networks.
\newblock {\em BMC Bioinform.}, 21(11):1--7, 2020.

\bibitem{haggenmuller2021skin}
Sarah Haggenm{\"u}ller, Roman~C Maron, Achim Hekler, Jochen~S Utikal, Catarina
  Barata, Raymond~L Barnhill, Helmut Beltraminelli, Carola Berking, Brigid
  Betz-Stablein, Andreas Blum, et~al.
\newblock Skin cancer classification via convolutional neural networks:
  systematic review of studies involving human experts.
\newblock {\em Eur. J. Cancer}, 156:202--216, 2021.

\bibitem{bitencourt2021ai}
Almir Bitencourt, Isaac~Daimiel Naranjo, Roberto~Lo Gullo, Carolina~Rossi
  Saccarelli, and Katja Pinker.
\newblock Ai-enhanced breast imaging: Where are we and where are we heading?
\newblock {\em Eur. J. Radiol.}, 142:109882, 2021.

\bibitem{duggento2021deep}
Andrea Duggento, Allegra Conti, Alessandro Mauriello, Maria Guerrisi, and
  Nicola Toschi.
\newblock Deep computational pathology in breast cancer.
\newblock In {\em Semin. Cancer Biol.}, volume~72, pages 226--237. Elsevier,
  2021.

\bibitem{michael2021breast}
Epimack Michael, He~Ma, Hong Li, Frank Kulwa, and Jing Li.
\newblock Breast cancer segmentation methods: current status and future
  potentials.
\newblock {\em Biomed Res. Int.}, 2021, 2021.

\bibitem{ramadan2020methods}
Saleem~Z Ramadan.
\newblock Methods used in computer-aided diagnosis for breast cancer detection
  using mammograms: a review.
\newblock {\em J. Healthc. Eng.}, 2020, 2020.

\bibitem{abdelhafiz2019deep}
Dina Abdelhafiz, Clifford Yang, Reda Ammar, and Sheida Nabavi.
\newblock Deep convolutional neural networks for mammography: advances,
  challenges and applications.
\newblock {\em BMC Bioinform.}, 20(11):1--20, 2019.

\bibitem{hosni2019reviewing}
Mohamed Hosni, Ibtissam Abnane, Ali Idri, Juan M~Carrillo de~Gea, and Jos{\'e}
  Luis~Fern{\'a}ndez Alem{\'a}n.
\newblock Reviewing ensemble classification methods in breast cancer.
\newblock {\em Comput. Meth. Prog. Bio.}, 177:89--112, 2019.

\bibitem{fujioka2020utility}
Tomoyuki Fujioka, Mio Mori, Kazunori Kubota, Jun Oyama, Emi Yamaga, Yuka
  Yashima, Leona Katsuta, Kyoko Nomura, Miyako Nara, Goshi Oda, et~al.
\newblock The utility of deep learning in breast ultrasonic imaging: a review.
\newblock {\em Diagnostics}, 10(12):1055, 2020.

\bibitem{ibrahim2020artificial}
Asmaa Ibrahim, Paul Gamble, Ronnachai Jaroensri, Mohammed~M Abdelsamea, Craig~H
  Mermel, Po-Hsuan~Cameron Chen, and Emad~A Rakha.
\newblock Artificial intelligence in digital breast pathology: techniques and
  applications.
\newblock {\em The Breast}, 49:267--273, 2020.

\bibitem{abdelrahman2021convolutional}
Leila Abdelrahman, Manal Al~Ghamdi, Fernando Collado-Mesa, and Mohamed
  Abdel-Mottaleb.
\newblock Convolutional neural networks for breast cancer detection in
  mammography: A survey.
\newblock {\em Comput. Biol. Med.}, 131:104248, 2021.

\bibitem{gardezi2019breast}
Syed Jamal~Safdar Gardezi, Ahmed Elazab, Baiying Lei, and Tianfu Wang.
\newblock Breast cancer detection and diagnosis using mammographic data:
  systematic review.
\newblock {\em J. Med. Internet Res.}, 21(7):e14464, 2019.

\bibitem{rashmi2022breast}
R~Rashmi, Keerthana Prasad, and Chethana Babu~K Udupa.
\newblock Breast histopathological image analysis using image processing
  techniques for diagnostic puposes: A methodological review.
\newblock {\em J. Med. Sys.}, 46(1):1--24, 2022.

\bibitem{ayana2021transfer}
Gelan Ayana, Kokeb Dese, and Se-woon Choe.
\newblock Transfer learning in breast cancer diagnoses via ultrasound imaging.
\newblock {\em Cancers}, 13(4):738, 2021.

\bibitem{debelee2020deep}
Taye~Girma Debelee, Samuel~Rahimeto Kebede, Friedhelm Schwenker, and
  Zemene~Matewos Shewarega.
\newblock Deep learning in selected cancers’ image analysis—a survey.
\newblock {\em J. Imaging}, 6(11):121, 2020.

\bibitem{ali2021state}
Saqib Ali, Jianqiang Li, Yan Pei, Rooha Khurram, Abdul~Basit Rasool, et~al.
\newblock State-of-the-art challenges and perspectives in multi-organ cancer
  diagnosis via deep learning-based methods.
\newblock {\em Cancers}, 13(21):5546, 2021.

\bibitem{lee2021axillary}
Yan-Wei Lee, Chiun-Sheng Huang, Chung-Chih Shih, and Ruey-Feng Chang.
\newblock Axillary lymph node metastasis status prediction of early-stage
  breast cancer using convolutional neural networks.
\newblock {\em Comput. Biol. Med.}, 130:104206, 2021.

\bibitem{ketabi2021computer}
Hossein Ketabi, Ali Ekhlasi, and Hessam Ahmadi.
\newblock A computer-aided approach for automatic detection of breast masses in
  digital mammogram via spectral clustering and support vector machine.
\newblock {\em Phys. Eng. Sci.}, 44(1):277--290, 2021.

\bibitem{oyelade2021bioinspired}
Olaide~N Oyelade and Absalom~E Ezugwu.
\newblock A bioinspired neural architecture search based convolutional neural
  network for breast cancer detection using histopathology images.
\newblock {\em Sci. Rep.}, 11(1):1--28, 2021.

\bibitem{singla2019automated}
Neeru Singla, Kavita Dubey, and Vishal Srivastava.
\newblock Automated assessment of breast cancer margin in optical coherence
  tomography images via pretrained convolutional neural network.
\newblock {\em J. Biophotonics}, 12(3):e201800255, 2019.

\bibitem{shen2021interpretable}
Yiqiu Shen, Nan Wu, Jason Phang, Jungkyu Park, Kangning Liu, Sudarshini Tyagi,
  Laura Heacock, S~Gene Kim, Linda Moy, Kyunghyun Cho, et~al.
\newblock An interpretable classifier for high-resolution breast cancer
  screening images utilizing weakly supervised localization.
\newblock {\em Med. Image Anal.}, 68:101908, 2021.

\bibitem{saffari2020fully}
Nasibeh Saffari, Hatem~A Rashwan, Mohamed Abdel-Nasser, Vivek Kumar~Singh,
  Meritxell Arenas, Eleni Mangina, Blas Herrera, and Domenec Puig.
\newblock Fully automated breast density segmentation and classification using
  deep learning.
\newblock {\em Diagnostics}, 10(11):988, 2020.

\bibitem{kimeswenger2021artificial}
Susanne Kimeswenger, Philipp Tschandl, Petar Noack, Markus Hofmarcher,
  Elisabeth Rumetshofer, Harald Kindermann, Rene Silye, Sepp Hochreiter, Martin
  Kaltenbrunner, Emmanuella Guenova, et~al.
\newblock Artificial neural networks and pathologists recognize basal cell
  carcinomas based on different histological patterns.
\newblock {\em Modern Pathol.}, 34(5):895--903, 2021.

\bibitem{cao2019experimental}
Zhantao Cao, Lixin Duan, Guowu Yang, Ting Yue, and Qin Chen.
\newblock An experimental study on breast lesion detection and classification
  from ultrasound images using deep learning architectures.
\newblock {\em BMC Med. Imaging}, 19(1):1--9, 2019.

\bibitem{jaber2020deep}
Mustafa~I Jaber, Bing Song, Clive Taylor, Charles~J Vaske, Stephen~C Benz,
  Shahrooz Rabizadeh, Patrick Soon-Shiong, and Christopher~W Szeto.
\newblock A deep learning image-based intrinsic molecular subtype classifier of
  breast tumors reveals tumor heterogeneity that may affect survival.
\newblock {\em Breast Cancer Res.}, 22(1):1--10, 2020.

\bibitem{george2020breast}
Kalpana George, Shameer Faziludeen, Praveen Sankaran, et~al.
\newblock Breast cancer detection from biopsy images using nucleus guided
  transfer learning and belief based fusion.
\newblock {\em Comput. Biol. Med.}, 124:103954, 2020.

\bibitem{chowdhary2020efficient}
Chiranji~Lal Chowdhary, Mohit Mittal, PA~Pattanaik, Zbigniew Marszalek, et~al.
\newblock An efficient segmentation and classification system in medical images
  using intuitionist possibilistic fuzzy c-mean clustering and fuzzy svm
  algorithm.
\newblock {\em Sensors}, 20(14):3903, 2020.

\bibitem{agnes2020appraisal}
S~Akila Agnes and J~Anitha.
\newblock Appraisal of deep-learning techniques on computer-aided lung cancer
  diagnosis with computed tomography screening.
\newblock {\em J. Med. Phys.}, 45(2):98, 2020.

\bibitem{chaudhury2021breast}
Sushovan Chaudhury, Manik Rakhra, Naz Memon, Kartik Sau, and Melkamu~Teshome
  Ayana.
\newblock Breast cancer calcifications: Identification using a novel
  segmentation approach.
\newblock {\em Comput. Math. Method M.}, 2021, 2021.

\bibitem{kassani2019comparative}
Sara~Hosseinzadeh Kassani and Peyman~Hosseinzadeh Kassani.
\newblock A comparative study of deep learning architectures on melanoma
  detection.
\newblock {\em Tissue Cell}, 58:76--83, 2019.

\bibitem{melekoodappattu2019hybridized}
Jayesh~George Melekoodappattu and Perumal~Sankar Subbian.
\newblock A hybridized elm for automatic micro calcification detection in
  mammogram images based on multi-scale features.
\newblock {\em J. Med. Sys.}, 43(7):1--12, 2019.

\bibitem{saba2019region}
Tanzila Saba, Muhammad~Attique Khan, Amjad Rehman, and Souad~Larabi
  Marie-Sainte.
\newblock Region extraction and classification of skin cancer: A heterogeneous
  framework of deep cnn features fusion and reduction.
\newblock {\em J. Med. Sys.}, 43(9):1--19, 2019.

\bibitem{wei2021optimized}
Liu Wei, Su~Xiao Pan, YA~Nanehkaran, and V~Rajinikanth.
\newblock An optimized method for skin cancer diagnosis using modified thermal
  exchange optimization algorithm.
\newblock {\em Comput. Math. Method M.}, 2021, 2021.

\bibitem{laang2021can}
Kristina L{\aa}ng, Solveig Hofvind, Alejandro Rodr{\'\i}guez-Ruiz, and Ingvar
  Andersson.
\newblock Can artificial intelligence reduce the interval cancer rate in
  mammography screening?
\newblock {\em Eur. Radiol.}, 31(8):5940--5947, 2021.

\bibitem{li2020multi}
Xuechen Li, Linlin Shen, Xinpeng Xie, Shiyun Huang, Zhien Xie, Xian Hong, and
  Juan Yu.
\newblock Multi-resolution convolutional networks for chest x-ray radiograph
  based lung nodule detection.
\newblock {\em Artif. Intell. Med.}, 103:101744, 2020.

\bibitem{ciga2021overcoming}
Ozan Ciga, Tony Xu, Sharon Nofech-Mozes, Shawna Noy, Fang-I Lu, and Anne~L
  Martel.
\newblock Overcoming the limitations of patch-based learning to detect cancer
  in whole slide images.
\newblock {\em Sci. Rep.}, 11(1):1--10, 2021.

\bibitem{george2020computer}
Kalpana George, Praveen Sankaran, et~al.
\newblock Computer assisted recognition of breast cancer in biopsy images via
  fusion of nucleus-guided deep convolutional features.
\newblock {\em Comput. Meth. Prog. Bio.}, 194:105531, 2020.

\bibitem{wang2019lung}
Qin Wang, Fengyi Shen, Linyao Shen, Jia Huang, and Weiguang Sheng.
\newblock Lung nodule detection in ct images using a raw patch-based
  convolutional neural network.
\newblock {\em J. Digit. Imaging}, 32(6):971--979, 2019.

\bibitem{jojoa2021melanoma}
Mario~Fernando Jojoa~Acosta, Liesle~Yail Caballero~Tovar, Maria~Begonya
  Garcia-Zapirain, and Winston~Spencer Percybrooks.
\newblock Melanoma diagnosis using deep learning techniques on dermatoscopic
  images.
\newblock {\em BMC Med. Imaging}, 21(1):1--11, 2021.

\bibitem{lu2021data}
Ming~Y Lu, Drew~FK Williamson, Tiffany~Y Chen, Richard~J Chen, Matteo Barbieri,
  and Faisal Mahmood.
\newblock Data-efficient and weakly supervised computational pathology on
  whole-slide images.
\newblock {\em Nat. Biomed.}, 5(6):555--570, 2021.

\bibitem{cai2019breast}
Hongmin Cai, Qinjian Huang, Wentao Rong, Yan Song, Jiao Li, Jinhua Wang,
  Jiazhou Chen, and Li~Li.
\newblock Breast microcalcification diagnosis using deep convolutional neural
  network from digital mammograms.
\newblock {\em Comput. Math. Method M.}, 2019, 2019.

\bibitem{liu2020automatic}
Lina Liu, Lichao Mou, Xiao~Xiang Zhu, and Mrinal Mandal.
\newblock Automatic skin lesion classification based on mid-level feature
  learning.
\newblock {\em Comput. Med. Imag. Graph.}, 84:101765, 2020.

\bibitem{yu2020mammographic}
Xiangchun Yu, Wei Pang, Qing Xu, and Miaomiao Liang.
\newblock Mammographic image classification with deep fusion learning.
\newblock {\em Sci. Rep.}, 10(1):1--11, 2020.

\bibitem{johny2021dynamic}
Anil Johny and KN~Madhusoodanan.
\newblock Dynamic learning rate in deep cnn model for metastasis detection and
  classification of histopathology images.
\newblock {\em Comput. Math. Method M.}, 2021, 2021.

\bibitem{cao2021application}
Xingmei Cao, Jeng-Shyang Pan, Zhengdi Wang, Zhonghai Sun, Anwar ul~Haq, Wenyu
  Deng, and Shuangyuan Yang.
\newblock Application of generated mask method based on mask r-cnn in
  classification and detection of melanoma.
\newblock {\em Comput. Meth. Prog. Bio.}, 207:106174, 2021.

\bibitem{singh2020imbalanced}
Rishav Singh, Tanveer Ahmed, Abhinav Kumar, Amit~Kumar Singh, Anil~Kumar
  Pandey, and Sanjay~Kumar Singh.
\newblock Imbalanced breast cancer classification using transfer learning.
\newblock {\em IEEE/ACM Trans. Comput. Biol. Bioinform.}, 18(1):83--93, 2020.

\bibitem{khan2019construction}
Muhammad~Attique Khan, Tallha Akram, Muhammad Sharif, Tanzila Saba, Kashif
  Javed, Ikram~Ullah Lali, Urcun~John Tanik, and Amjad Rehman.
\newblock Construction of saliency map and hybrid set of features for efficient
  segmentation and classification of skin lesion.
\newblock {\em J. Med. Sys.}, 82(6):741--763, 2019.

\bibitem{wang2021false}
Bin Wang, Shuaizong Si, Hai Zhao, Hongbo Zhu, and Shengchang Dou.
\newblock False positive reduction in pulmonary nodule classification using 3d
  texture and edge feature in ct images.
\newblock {\em Technol. Health Care}, 29(6):1071--1088, 2021.

\bibitem{duggento2019ad}
Andrea Duggento, Marco Aiello, Carlo Cavaliere, Giuseppe~L Cascella, Davide
  Cascella, Giovanni Conte, Maria Guerrisi, and Nicola Toschi.
\newblock An ad hoc random initialization deep neural network architecture for
  discriminating malignant breast cancer lesions in mammographic images.
\newblock {\em Contrast Med. Mol. I.}, 2019, 2019.

\bibitem{yuan2021efficient}
Haiying Yuan, Zhongwei Fan, Yanrui Wu, and Junpeng Cheng.
\newblock An efficient multi-path 3d convolutional neural network for
  false-positive reduction of pulmonary nodule detection.
\newblock {\em Int. J. Comput. Ass. Rad.}, pages 1--9, 2021.

\bibitem{zhang2021improving}
Yu-Dong Zhang, Suresh~Chandra Satapathy, Di~Wu, David~S Guttery, Juan~Manuel
  G{\'o}rriz, and Shui-Hua Wang.
\newblock Improving ductal carcinoma in situ classification by convolutional
  neural network with exponential linear unit and rank-based weighted pooling.
\newblock {\em Complex Intell. Syst.}, 7(3):1295--1310, 2021.

\bibitem{omigbodun2019effects}
Akinyinka~O Omigbodun, Frederic Noo, Michael McNitt-Gray, William Hsu, and
  Scott~S Hsieh.
\newblock The effects of physics-based data augmentation on the
  generalizability of deep neural networks: Demonstration on nodule
  false-positive reduction.
\newblock {\em Med. Phys.}, 46(10):4563--4574, 2019.

\bibitem{mutepfe2021generative}
Freedom Mutepfe, Behnam~Kiani Kalejahi, Saeed Meshgini, and Sebelan Danishvar.
\newblock Generative adversarial network image synthesis method for skin lesion
  generation and classification.
\newblock {\em J. Med. Signals Sens.}, 11(4):237, 2021.

\bibitem{pang2021semi}
Ting Pang, Jeannie Hsiu~Ding Wong, Wei~Lin Ng, and Chee~Seng Chan.
\newblock Semi-supervised gan-based radiomics model for data augmentation in
  breast ultrasound mass classification.
\newblock {\em Comput. Meth. Prog. Bio.}, 203:106018, 2021.

\bibitem{rashid2019skin}
Haroon Rashid, M~Asjid Tanveer, and Hassan~Aqeel Khan.
\newblock Skin lesion classification using gan based data augmentation.
\newblock In {\em 2019 41st Annual International Conference of the IEEE
  Engineering in Medicine and Biology Society (EMBC)}, pages 916--919. IEEE,
  2019.

\bibitem{ahmad2021improving}
Bilal Ahmad, Sun Jun, Vasile Palade, Qi~You, Li~Mao, and Mao Zhongjie.
\newblock Improving skin cancer classification using heavy-tailed student
  t-distribution in generative adversarial networks (ted-gan).
\newblock {\em Diagnostics}, 11(11):2147, 2021.

\bibitem{ramadan2020using}
Saleem~Z Ramadan.
\newblock Using convolutional neural network with cheat sheet and data
  augmentation to detect breast cancer in mammograms.
\newblock {\em Comput. Math. Method M.}, 2020, 2020.

\bibitem{sadad2021internet}
Tariq Sadad, Amjad~Rehman Khan, Ayyaz Hussain, Usman Tariq, Suliman~Mohamed
  Fati, Saeed~Ali Bahaj, and Asim Munir.
\newblock Internet of medical things embedding deep learning with data
  augmentation for mammogram density classification.
\newblock {\em Microsc. Res. Techniq.}, 2021.

\bibitem{wang2021boosted}
Jun Wang, Qianying Liu, Haotian Xie, Zhaogang Yang, and Hefeng Zhou.
\newblock Boosted efficientnet: detection of lymph node metastases in breast
  cancer using convolutional neural networks.
\newblock {\em Cancers}, 13(4):661, 2021.

\bibitem{wang2021automatic}
Bowen Wang, Toshihiro Takeda, Kento Sugimoto, Jiahao Zhang, Shoya Wada, Shozo
  Konishi, Shirou Manabe, Katsuki Okada, and Yasushi Matsumura.
\newblock Automatic creation of annotations for chest radiographs based on the
  positional information extracted from radiographic image reports.
\newblock {\em Comput. Meth. Prog. Bio.}, 209:106331, 2021.

\bibitem{lotter2021robust}
William Lotter, Abdul~Rahman Diab, Bryan Haslam, Jiye~G Kim, Giorgia Grisot,
  Eric Wu, Kevin Wu, Jorge~Onieva Onieva, Yun Boyer, Jerrold~L Boxerman, et~al.
\newblock Robust breast cancer detection in mammography and digital breast
  tomosynthesis using an annotation-efficient deep learning approach.
\newblock {\em Nat. Med.}, 27(2):244--249, 2021.

\bibitem{qi2018label}
Qi~Qi, Yanlong Li, Jitian Wang, Han Zheng, Yue Huang, Xinghao Ding, and
  Gustavo~Kunde Rohde.
\newblock Label-efficient breast cancer histopathological image classification.
\newblock {\em IEEE J. Biomed. Health Inform.}, 23(5):2108--2116, 2018.

\bibitem{yusoff2020somatic}
Abdul Aziz~Mohamed Yusoff, Siti Zulaikha Nashwa~Mohd Khair, Wan Salihah~Wan
  Abdullah, Siti~Muslihah Abd~Radzak, Jafri~Malin Abdullah, et~al.
\newblock Somatic mitochondrial dna d-loop mutations in meningioma discovered:
  a preliminary data.
\newblock {\em J. Cancer Res. Ther.}, 16(6):1517, 2020.

\bibitem{gu2019automatic}
Yu~Gu, Xiaoqi Lu, Baohua Zhang, Ying Zhao, Dahua Yu, Lixin Gao, Guimei Cui,
  Liang Wu, and Tao Zhou.
\newblock Automatic lung nodule detection using multi-scale dot
  nodule-enhancement filter and weighted support vector machines in chest
  computed tomography.
\newblock {\em PLoS one}, 14(1):e0210551, 2019.

\bibitem{irfan2021dilated}
Rizwana Irfan, Abdulwahab~Ali Almazroi, Hafiz~Tayyab Rauf, Robertas
  Dama{\v{s}}evi{\v{c}}ius, Emad~Abouel Nasr, and Abdelatty~E Abdelgawad.
\newblock Dilated semantic segmentation for breast ultrasonic lesion detection
  using parallel feature fusion.
\newblock {\em Diagnostics}, 11(7):1212, 2021.

\bibitem{kofler2021robust}
Florian Kofler, Ivan Ezhov, Lucas Fidon, Carolin~M Pirkl, Johannes~C Paetzold,
  Egon Burian, Sarthak Pati, Malek El~Husseini, Fernando Navarro, Suprosanna
  Shit, et~al.
\newblock Robust, primitive, and unsupervised quality estimation for
  segmentation ensembles.
\newblock {\em Front. Neurosci.}, 15, 2021.

\bibitem{ma2019automated}
Xiangyuan Ma, Jun Wei, Chuan Zhou, Mark~A Helvie, Heang-Ping Chan, Lubomir~M
  Hadjiiski, and Yao Lu.
\newblock Automated pectoral muscle identification on mlo-view mammograms:
  Comparison of deep neural network to conventional computer vision.
\newblock {\em Med. Phys.}, 46(5):2103--2114, 2019.

\bibitem{lee2021comparison}
Rebecca~Sawyer Lee, Jared~A Dunnmon, Ann He, Siyi Tang, Christopher Re, and
  Daniel~L Rubin.
\newblock Comparison of segmentation-free and segmentation-dependent
  computer-aided diagnosis of breast masses on a public mammography dataset.
\newblock {\em J. Biomed. Inform.}, 113:103656, 2021.

\bibitem{yu2021design}
Hui Yu, Jinqiu Li, Lixin Zhang, Yuzhen Cao, Xuyao Yu, and Jinglai Sun.
\newblock Design of lung nodules segmentation and recognition algorithm based
  on deep learning.
\newblock {\em BMC Bioinform.}, 22(5):1--21, 2021.

\bibitem{nasrullah2019automated}
Nasrullah Nasrullah, Jun Sang, Mohammad~S Alam, Muhammad Mateen, Bin Cai, and
  Haibo Hu.
\newblock Automated lung nodule detection and classification using deep
  learning combined with multiple strategies.
\newblock {\em Sensors}, 19(17):3722, 2019.

\bibitem{makrogiannis2021discriminative}
Sokratis Makrogiannis, Keni Zheng, and Chelsea Harris.
\newblock Discriminative localized sparse approximations for mass
  characterization in mammograms.
\newblock {\em Front. Oncol.}, 11:725320--725320, 2021.

\bibitem{li2021dual}
Heyi Li, Dongdong Chen, William~H Nailon, Mike~E Davies, and David~I Laurenson.
\newblock Dual convolutional neural networks for breast mass segmentation and
  diagnosis in mammography.
\newblock {\em IEEE Trans. Med. Imaging}, 41(1):3--13, 2021.

\bibitem{garcia2019segmentation}
Jose~Luis Garcia-Arroyo and Begonya Garcia-Zapirain.
\newblock Segmentation of skin lesions in dermoscopy images using fuzzy
  classification of pixels and histogram thresholding.
\newblock {\em Comput. Meth. Prog. Bio.}, 168:11--19, 2019.

\bibitem{mahbod2020effects}
Amirreza Mahbod, Philipp Tschandl, Georg Langs, Rupert Ecker, and Isabella
  Ellinger.
\newblock The effects of skin lesion segmentation on the performance of
  dermatoscopic image classification.
\newblock {\em Comput. Meth. Prog. Bio.}, 197:105725, 2020.

\bibitem{khan2019effective}
Sajid~Ali Khan, Shariq Hussain, Shunkun Yang, and Khalid Iqbal.
\newblock Effective and reliable framework for lung nodules detection from ct
  scan images.
\newblock {\em Sci. Rep.}, 9(1):1--14, 2019.

\bibitem{yang2019comprehensive}
Wenming Yang, Zirui Wang, Kaiquan Chen, Zhide Li, and Qingmin Liao.
\newblock A comprehensive classification system for breast cancer diagnosis
  based on dynamic optical breast imaging.
\newblock In {\em 2019 41st Annual International Conference of the IEEE
  Engineering in Medicine and Biology Society (EMBC)}, pages 4741--4744. IEEE,
  2019.

\bibitem{kumar2021classification}
Vinod Kumar and Brijesh Bakariya.
\newblock Classification of malignant lung cancer using deep learning.
\newblock {\em J. Med. Eng. Technol.}, 45(2):85--93, 2021.

\bibitem{zanddizari2021new}
Hadi Zanddizari, Nam Nguyen, Behnam Zeinali, and J~Morris Chang.
\newblock A new preprocessing approach to improve the performance of cnn-based
  skin lesion classification.
\newblock {\em Med. Biol. Eng. Comput.}, 59(5):1123--1131, 2021.

\bibitem{sarath2020two}
Chandra~K Sarath, Arunava Chakravarty, Nirmalya Ghosh, Tandra Sarkar,
  Ramanathan Sethuraman, and Debdoot Sheet.
\newblock A two-stage multiple instance learning framework for the detection of
  breast cancer in mammograms.
\newblock In {\em 2020 42nd Annual International Conference of the IEEE
  Engineering in Medicine \& Biology Society (EMBC)}, pages 1128--1131. IEEE,
  2020.

\bibitem{saba2019lung}
Tanzila Saba, Ahmed Sameh, Fatima Khan, Shafqat~Ali Shad, and Muhammad Sharif.
\newblock Lung nodule detection based on ensemble of hand crafted and deep
  features.
\newblock {\em J. Med. Sys.}, 43(12):1--12, 2019.

\bibitem{kavitha2021deep}
T~Kavitha, Paul~P Mathai, C~Karthikeyan, M~Ashok, Rachna Kohar, J~Avanija, and
  S~Neelakandan.
\newblock Deep learning based capsule neural network model for breast cancer
  diagnosis using mammogram images.
\newblock {\em Interdiscip. Sci. Comput. Life Sci.}, pages 1--17, 2021.

\bibitem{piantadosi2020multi}
Gabriele Piantadosi, Mario Sansone, Roberta Fusco, and Carlo Sansone.
\newblock Multi-planar 3d breast segmentation in mri via deep convolutional
  neural networks.
\newblock {\em Artif. Intell. Med.}, 103:101781, 2020.

\bibitem{blanc2020artificial}
D~Blanc, V~Racine, A~Khalil, M~Deloche, J-A Broyelle, I~Hammouamri,
  E~Sinitambirivoutin, M~Fiammante, E~Verdier, T~Besson, et~al.
\newblock Artificial intelligence solution to classify pulmonary nodules on ct.
\newblock {\em Diagn. Interv. Imag.}, 101(12):803--810, 2020.

\bibitem{lu2021breast}
Xi~Lu, Zejun You, Miaomiao Sun, Jing Wu, and Zhihong Zhang.
\newblock Breast cancer mitotic cell detection using cascade convolutional
  neural network with u-net.
\newblock {\em Math. Biosci. Eng}, 18:673--695, 2021.

\bibitem{jimenez2019deep}
Gabriel Jim{\'e}nez and Daniel Racoceanu.
\newblock Deep learning for semantic segmentation vs. classification in
  computational pathology: application to mitosis analysis in breast cancer
  grading.
\newblock {\em Front. Bioeng. Biotechnol.}, 7:145, 2019.

\bibitem{alom2019recurrent}
Md~Zahangir Alom, Chris Yakopcic, Mahmudul Hasan, Tarek~M Taha, and Vijayan~K
  Asari.
\newblock Recurrent residual u-net for medical image segmentation.
\newblock {\em J. Med. Imaging}, 6(1):014006, 2019.

\bibitem{kim2021computer}
Chan-Il Kim, Seok-Min Hwang, Eun-Bin Park, Chang-Hee Won, and Jong-Ha Lee.
\newblock Computer-aided diagnosis algorithm for classification of malignant
  melanoma using deep neural networks.
\newblock {\em Sensors}, 21(16):5551, 2021.

\bibitem{shen2020multicontext}
Rongbo Shen, Ke~Zhou, Kezhou Yan, Kuan Tian, and Jun Zhang.
\newblock Multicontext multitask learning networks for mass detection in
  mammogram.
\newblock {\em Med. Phys.}, 47(4):1566--1578, 2020.

\bibitem{agarwal2020deep}
Richa Agarwal, Oliver D{\'\i}az, Moi~Hoon Yap, Xavier Llado, and Robert Marti.
\newblock Deep learning for mass detection in full field digital mammograms.
\newblock {\em Comput. Biol. Med.}, 121:103774, 2020.

\bibitem{pourasad2021presentation}
Yaghoub Pourasad, Esmaeil Zarouri, Mohammad Salemizadeh~Parizi, and Amin
  Salih~Mohammed.
\newblock Presentation of novel architecture for diagnosis and identifying
  breast cancer location based on ultrasound images using machine learning.
\newblock {\em Diagnostics}, 11(10):1870, 2021.

\bibitem{sasikala2019particle}
S~Sasikala, M~Bharathi, M~Ezhilarasi, Sathiya Senthil, and M~Ramasubba Reddy.
\newblock Particle swarm optimization based fusion of ultrasound echographic
  and elastographic texture features for improved breast cancer detection.
\newblock {\em Australas. Phys. Eng. Sci. Med.}, 42(3):677--688, 2019.

\bibitem{chang2020svm}
Chuan-Yu Chang, Kathiravan Srinivasan, Mao-Cheng Chen, and Shao-Jer Chen.
\newblock Svm-enabled intelligent genetic algorithmic model for realizing
  efficient universal feature selection in breast cyst image acquired via
  ultrasound sensing systems.
\newblock {\em Sensors}, 20(2):432, 2020.

\bibitem{kailasam2019novel}
S~Piramu Kailasam and M~Mohamed Sathik.
\newblock A novel hybrid feature extraction model for classification on
  pulmonary nodules.
\newblock {\em Asian Pac. J. Cancer Prev.}, 20(2):457, 2019.

\bibitem{chatterjee2019integration}
Saptarshi Chatterjee, Debangshu Dey, and Sugata Munshi.
\newblock Integration of morphological preprocessing and fractal based feature
  extraction with recursive feature elimination for skin lesion types
  classification.
\newblock {\em Comput. Meth. Prog. Bio.}, 178:201--218, 2019.

\bibitem{li2019predicting}
Shulong Li, Panpan Xu, Bin Li, Liyuan Chen, Zhiguo Zhou, Hongxia Hao, Yingying
  Duan, Michael Folkert, Jianhua Ma, Shiying Huang, et~al.
\newblock Predicting lung nodule malignancies by combining deep convolutional
  neural network and handcrafted features.
\newblock {\em Phys. Med. Biol.}, 64(17):175012, 2019.

\bibitem{sakai2020method}
Ayaka Sakai, Yuya Onishi, Misaki Matsui, Hidetoshi Adachi, Atsushi Teramoto,
  Kuniaki Saito, and Hiroshi Fujita.
\newblock A method for the automated classification of benign and malignant
  masses on digital breast tomosynthesis images using machine learning and
  radiomic features.
\newblock {\em Radiol. Phys. Technol.}, 13(1):27--36, 2020.

\bibitem{lei2019mammography}
Chuqian Lei, Wei Wei, Zhenyu Liu, Qianqian Xiong, Ciqiu Yang, Mei Yang, Liulu
  Zhang, Teng Zhu, Xiaosheng Zhuang, Chunling Liu, et~al.
\newblock Mammography-based radiomic analysis for predicting benign bi-rads
  category 4 calcifications.
\newblock {\em Eur. J. Radiol.}, 121:108711, 2019.

\bibitem{alsaade2021developing}
Fawaz~Waselallah Alsaade, Theyazn~HH Aldhyani, and Mosleh~Hmoud Al-Adhaileh.
\newblock Developing a recognition system for diagnosing melanoma skin lesions
  using artificial intelligence algorithms.
\newblock {\em Comput. Math. Method M.}, 2021, 2021.

\bibitem{manzo2020bucket}
Mario Manzo and Simone Pellino.
\newblock Bucket of deep transfer learning features and classification models
  for melanoma detection.
\newblock {\em J. Imaging}, 6(12):129, 2020.

\bibitem{ghazouani2021towards}
Haythem Ghazouani and Walid Barhoumi.
\newblock Towards non-data-hungry and fully-automated diagnosis of breast
  cancer from mammographic images.
\newblock {\em Comput. Biol. Med.}, 139:105011, 2021.

\bibitem{ozdemir20193d}
Onur Ozdemir, Rebecca~L Russell, and Andrew~A Berlin.
\newblock A 3d probabilistic deep learning system for detection and diagnosis
  of lung cancer using low-dose ct scans.
\newblock {\em IEEE Trans. Med. Imaging}, 39(5):1419--1429, 2019.

\bibitem{rashmi2021bchisto}
R~Rashmi, Keerthana Prasad, and Chethana Babu~K Udupa.
\newblock Bchisto-net: Breast histopathological image classification by global
  and local feature aggregation.
\newblock {\em Artif. Intell. Med.}, 121:102191, 2021.

\bibitem{ameri2020deep}
A~Ameri.
\newblock A deep learning approach to skin cancer detection in dermoscopy
  images.
\newblock {\em J. Biomed. Phys. Eng.}, 10(6):801, 2020.

\bibitem{heenaye2021multi}
Maleika Heenaye-Mamode~Khan, Nazmeen Boodoo-Jahangeer, Wasiimah Dullull,
  Shaista Nathire, Xiaohong Gao, GR~Sinha, and Kapil~Kumar Nagwanshi.
\newblock Multi-class classification of breast cancer abnormalities using deep
  convolutional neural network (cnn).
\newblock {\em PLoS one}, 16(8):e0256500, 2021.

\bibitem{shu2020deep}
Xin Shu, Lei Zhang, Zizhou Wang, Qing Lv, and Zhang Yi.
\newblock Deep neural networks with region-based pooling structures for
  mammographic image classification.
\newblock {\em IEEE Trans. Med. Imaging}, 39(6):2246--2255, 2020.

\bibitem{chen2019computer}
Chia-Hung Chen, Yan-Wei Lee, Yao-Sian Huang, Wei-Ren Lan, Ruey-Feng Chang,
  Chih-Yen Tu, Chih-Yu Chen, and Wei-Chih Liao.
\newblock Computer-aided diagnosis of endobronchial ultrasound images using
  convolutional neural network.
\newblock {\em Comput. Meth. Prog. Bio.}, 177:175--182, 2019.

\bibitem{sharma2020conventional}
Shallu Sharma and Rajesh Mehra.
\newblock Conventional machine learning and deep learning approach for
  multi-classification of breast cancer histopathology images—a comparative
  insight.
\newblock {\em J. Digit. imaging}, 33(3):632--654, 2020.

\bibitem{zhang2019investigation}
Shu Zhang, Fangfang Han, Zhengrong Liang, Jiaxing Tan, Weiguo Cao, Yongfeng
  Gao, Marc Pomeroy, Kenneth Ng, and Wei Hou.
\newblock An investigation of cnn models for differentiating malignant from
  benign lesions using small pathologically proven datasets.
\newblock {\em Comput. Med. Imag. Graph.}, 77:101645, 2019.

\bibitem{daoud2020breast}
Mohammad~I Daoud, Samir Abdel-Rahman, Tariq~M Bdair, Mahasen~S Al-Najar,
  Feras~H Al-Hawari, and Rami Alazrai.
\newblock Breast tumor classification in ultrasound images using combined deep
  and handcrafted features.
\newblock {\em Sensors}, 20(23):6838, 2020.

\bibitem{mastouri2021bilinear}
Rekka Mastouri, Nawres Khlifa, Henda Neji, and Saoussen Hantous-Zannad.
\newblock A bilinear convolutional neural network for lung nodules
  classification on ct images.
\newblock {\em Int. J. Comput. Ass. Rad.}, 16(1):91--101, 2021.

\bibitem{hagerty2019deep}
Jason~R Hagerty, R~Joe Stanley, Haidar~A Almubarak, Norsang Lama, Reda Kasmi,
  Peng Guo, Rhett~J Drugge, Harold~S Rabinovitz, Margaret Oliviero, and
  William~V Stoecker.
\newblock Deep learning and handcrafted method fusion: higher diagnostic
  accuracy for melanoma dermoscopy images.
\newblock {\em IEEE J. Biomed. Health Inform.}, 23(4):1385--1391, 2019.

\bibitem{ragab2021framework}
Dina~A Ragab, Omneya Attallah, Maha Sharkas, Jinchang Ren, and Stephen
  Marshall.
\newblock A framework for breast cancer classification using multi-dcnns.
\newblock {\em Comput. Biol. Med.}, 131:104245, 2021.

\bibitem{zhuang2021breast}
Zhemin Zhuang, Zengbiao Yang, Alex Noel~Joseph Raj, Chuliang Wei, Pengcheng
  Jin, and Shuxin Zhuang.
\newblock Breast ultrasound tumor image classification using image
  decomposition and fusion based on adaptive multi-model spatial feature
  fusion.
\newblock {\em Comput. Meth. Prog. Bio.}, 208:106221, 2021.

\bibitem{uthoff2019machine}
Johanna Uthoff, Matthew~J Stephens, John~D Newell~Jr, Eric~A Hoffman, Jared
  Larson, Nicholas Koehn, Frank~A De~Stefano, Chrissy~M Lusk, Angela~S
  Wenzlaff, Donovan Watza, et~al.
\newblock Machine learning approach for distinguishing malignant and benign
  lung nodules utilizing standardized perinodular parenchymal features from ct.
\newblock {\em Med. Phys.}, 46(7):3207--3216, 2019.

\bibitem{ningrum2021deep}
Dina Nur~Anggraini Ningrum, Sheng-Po Yuan, Woon-Man Kung, Chieh-Chen Wu,
  I-Shiang Tzeng, Chu-Ya Huang, Jack Yu-Chuan Li, and Yao-Chin Wang.
\newblock Deep learning classifier with patient’s metadata of dermoscopic
  images in malignant melanoma detection.
\newblock {\em J. Multidiscip. Healthc.}, 14:877, 2021.

\bibitem{yin2021prediction}
Guotao Yin, Yingchao Song, Xiaofeng Li, Lei Zhu, Qian Su, Dong Dai, and Wengui
  Xu.
\newblock Prediction of mediastinal lymph node metastasis based on 18f-fdg
  pet/ct imaging using support vector machine in non-small cell lung cancer.
\newblock {\em Eur. Radiol.}, 31(6):3983--3992, 2021.

\bibitem{mahbod2019fusing}
Amirreza Mahbod, Gerald Schaefer, Isabella Ellinger, Rupert Ecker, Alain
  Pitiot, and Chunliang Wang.
\newblock Fusing fine-tuned deep features for skin lesion classification.
\newblock {\em Comput. Med. Imag. Graph.}, 71:19--29, 2019.

\bibitem{cheraghlou2021machine}
Shayan Cheraghlou, Praneeth Sadda, George~O Agogo, and Michael Girardi.
\newblock A machine-learning modified cart algorithm informs merkel cell
  carcinoma prognosis.
\newblock {\em Aust. J. Dermatol.}, 62(3):323--330, 2021.

\bibitem{areiza2020novel}
Henry~Jho{\'a}n Areiza-Laverde, Andr{\'e}s~Eduardo Castro-Ospina,
  Mar{\'\i}a~Liliana Hern{\'a}ndez, and Gloria~M D{\'\i}az.
\newblock A novel method for objective selection of information sources using
  multi-kernel svm and local scaling.
\newblock {\em Sensors}, 20(14):3919, 2020.

\bibitem{rana2019machine}
Soumya~Prakash Rana, Maitreyee Dey, Gianluigi Tiberi, Lorenzo Sani, Alessandro
  Vispa, Giovanni Raspa, Michele Duranti, Mohammad Ghavami, and Sandra Dudley.
\newblock Machine learning approaches for automated lesion detection in
  microwave breast imaging clinical data.
\newblock {\em Sci. Rep.}, 9(1):1--12, 2019.

\bibitem{aswathy2021svm}
MA~Aswathy and M~Jagannath.
\newblock An svm approach towards breast cancer classification from
  h\&e-stained histopathology images based on integrated features.
\newblock {\em Med. Biol. Eng. Comput.}, 59(9):1773--1783, 2021.

\bibitem{ding2021optical}
Ming Ding, Shi-yu Pan, Jing Huang, Cheng Yuan, Qiang Zhang, Xiao-li Zhu, and
  Yan Cai.
\newblock Optical coherence tomography for identification of malignant
  pulmonary nodules based on random forest machine learning algorithm.
\newblock {\em PLoS one}, 16(12):e0260600, 2021.

\bibitem{pesce2019learning}
Emanuele Pesce, Samuel~Joseph Withey, Petros-Pavlos Ypsilantis, Robert
  Bakewell, Vicky Goh, and Giovanni Montana.
\newblock Learning to detect chest radiographs containing pulmonary lesions
  using visual attention networks.
\newblock {\em Med. Image Anal.}, 53:26--38, 2019.

\bibitem{zunair2020melanoma}
Hasib Zunair and A~Ben Hamza.
\newblock Melanoma detection using adversarial training and deep transfer
  learning.
\newblock {\em Phys. Med. Biol.}, 65(13):135005, 2020.

\bibitem{xie2019semi}
Yutong Xie, Jianpeng Zhang, and Yong Xia.
\newblock Semi-supervised adversarial model for benign--malignant lung nodule
  classification on chest ct.
\newblock {\em Med. Image Anal.}, 57:237--248, 2019.

\bibitem{hosni2020mapping}
Mohamed Hosni, Gin{\'e}s Garc{\'\i}a-Mateos, Juan~M Carrillo-de Gea, Ali Idri,
  and Jos{\'e}~Luis Fern{\'a}ndez-Alem{\'a}n.
\newblock A mapping study of ensemble classification methods in lung cancer
  decision support systems.
\newblock {\em Med. Biol. Eng. Comput.}, 58(10):2177--2193, 2020.

\bibitem{zhang2019optimization}
Long Zhang, Hong~Jie Gao, Jianhua Zhang, and Benjamin Badami.
\newblock Optimization of the convolutional neural networks for automatic
  detection of skin cancer.
\newblock {\em Open Med.}, 15(1):27--37, 2019.

\bibitem{wang2021wdccnet}
Yan Wang, Zizhou Wang, Yangqin Feng, and Lei Zhang.
\newblock Wdccnet: Weighted double-classifier constraint neural network for
  mammographic image classification.
\newblock {\em IEEE Trans. Med. Imaging}, 2021.

\bibitem{tran2019improving}
Giang~Son Tran, Thi~Phuong Nghiem, Van~Thi Nguyen, Chi~Mai Luong, and
  Jean-Christophe Burie.
\newblock Improving accuracy of lung nodule classification using deep learning
  with focal loss.
\newblock {\em J. Healthc. Eng.}, 2019, 2019.

\bibitem{xie2018knowledge}
Yutong Xie, Yong Xia, Jianpeng Zhang, Yang Song, Dagan Feng, Michael Fulham,
  and Weidong Cai.
\newblock Knowledge-based collaborative deep learning for benign-malignant lung
  nodule classification on chest ct.
\newblock {\em IEEE Trans. Med. Imaging}, 38(4):991--1004, 2018.

\bibitem{rv2021augmenting}
Aswiga RV, Shanthi AP, et~al.
\newblock Augmenting transfer learning with feature extraction techniques for
  limited breast imaging datasets.
\newblock {\em J. Digit. Imaging}, 34(3):618--629, 2021.

\bibitem{zhang2020new}
Chen Zhang, Jumin Zhao, Jing Niu, and Dengao Li.
\newblock New convolutional neural network model for screening and diagnosis of
  mammograms.
\newblock {\em PLoS One}, 15(8):e0237674, 2020.

\bibitem{suresh2019novel}
Annamalai Suresh, R~Udendhran, M~Balamurgan, and R~Varatharajan.
\newblock A novel internet of things framework integrated with real time
  monitoring for intelligent healthcare environment.
\newblock {\em J. Med. Sys.}, 43(6):1--10, 2019.

\bibitem{gerard2018fissurenet}
Sarah~E Gerard, Taylor~J Patton, Gary~E Christensen, John~E Bayouth, and
  Joseph~M Reinhardt.
\newblock Fissurenet: a deep learning approach for pulmonary fissure detection
  in ct images.
\newblock {\em IEEE Trans. Med. Imag.}, 38(1):156--166, 2018.

\bibitem{yang2021momminet}
Zhicheng Yang, Zhenjie Cao, Yanbo Zhang, Yuxing Tang, Xiaohui Lin, Rushan
  Ouyang, Mingxiang Wu, Mei Han, Jing Xiao, Lingyun Huang, et~al.
\newblock Momminet-v2: Mammographic multi-view mass identification networks.
\newblock {\em Med. Image Anal.}, 73:102204, 2021.

\bibitem{montaha2021breastnet18}
Sidratul Montaha, Sami Azam, Abul Kalam Muhammad Rakibul~Haque Rafid, Pronab
  Ghosh, Md~Hasan, Mirjam Jonkman, Friso~De Boer, et~al.
\newblock Breastnet18: A high accuracy fine-tuned vgg16 model evaluated using
  ablation study for diagnosing breast cancer from enhanced mammography images.
\newblock {\em Biology}, 10(12):1347, 2021.

\bibitem{gu2019progressive}
Yanyang Gu, Zongyuan Ge, C~Paul Bonnington, and Jun Zhou.
\newblock Progressive transfer learning and adversarial domain adaptation for
  cross-domain skin disease classification.
\newblock {\em IEEE J. Biomed. Health Inform.}, 24(5):1379--1393, 2019.

\bibitem{traore2020evaluating}
Abdarahmane Traor{\'e}, Abdoulaye~O Ly, and Moulay~A Akhloufi.
\newblock Evaluating deep learning algorithms in pulmonary nodule detection.
\newblock In {\em 2020 42nd Annual International Conference of the IEEE
  Engineering in Medicine \& Biology Society (EMBC)}, pages 1335--1338. IEEE,
  2020.

\bibitem{abdolahi2020artificial}
Mohammad Abdolahi, Mohammad Salehi, Iman Shokatian, and Reza Reiazi.
\newblock Artificial intelligence in automatic classification of invasive
  ductal carcinoma breast cancer in digital pathology images.
\newblock {\em Med. J. Islam Repub. Iran}, 34:140, 2020.

\bibitem{hasan2021comparative}
Mohammed~Rakeibul Hasan, Mohammed~Ishraaf Fatemi, Mohammad Monirujjaman~Khan,
  Manjit Kaur, and Atef Zaguia.
\newblock Comparative analysis of skin cancer (benign vs. malignant) detection
  using convolutional neural networks.
\newblock {\em J. Healthc. Eng.}, 2021, 2021.

\bibitem{singhal2020comparing}
Aditi Singhal, Ramesht Shukla, Pavan~Kumar Kankar, Saurabh Dubey, Sukhjeet
  Singh, and Ram~Bilas Pachori.
\newblock Comparing the capabilities of transfer learning models to detect skin
  lesion in humans.
\newblock {\em Proc. Inst. Mech. Eng. H: J. Eng. Med.}, 234(10):1083--1093,
  2020.

\bibitem{yu2020reproducible}
Kun-Hsing Yu, Tsung-Lu~Michael Lee, Ming-Hsuan Yen, SC~Kou, Bruce Rosen,
  Jung-Hsien Chiang, Isaac~S Kohane, et~al.
\newblock Reproducible machine learning methods for lung cancer detection using
  computed tomography images: Algorithm development and validation.
\newblock {\em J. Med. Internet Res.}, 22(8):e16709, 2020.

\bibitem{jain2021deep}
Satin Jain, Udit Singhania, Balakrushna Tripathy, Emad~Abouel Nasr, Mohamed~K
  Aboudaif, and Ali~K Kamrani.
\newblock Deep learning-based transfer learning for classification of skin
  cancer.
\newblock {\em Sensors}, 21(23):8142, 2021.

\bibitem{jaworek2021interpretability}
Joanna Jaworek-Korjakowska, Andrzej Brodzicki, Bill Cassidy, Connah Kendrick,
  and Moi~Hoon Yap.
\newblock Interpretability of a deep learning based approach for the
  classification of skin lesions into main anatomic body sites.
\newblock {\em Cancers}, 13(23):6048, 2021.

\bibitem{attallah2021intelligent}
Omneya Attallah and Maha Sharkas.
\newblock Intelligent dermatologist tool for classifying multiple skin cancer
  subtypes by incorporating manifold radiomics features categories.
\newblock {\em Contrast Media Mol. I.}, 2021, 2021.

\bibitem{tanaka2019computer}
Hiroki Tanaka, Shih-Wei Chiu, Takanori Watanabe, Setsuko Kaoku, and Takuhiro
  Yamaguchi.
\newblock Computer-aided diagnosis system for breast ultrasound images using
  deep learning.
\newblock {\em Phys. Med. Biol.}, 64(23):235013, 2019.

\bibitem{alzahrani2021comprehensive}
Saeed Alzahrani, Baidaa Al-Bander, and Waleed Al-Nuaimy.
\newblock A comprehensive evaluation and benchmarking of convolutional neural
  networks for melanoma diagnosis.
\newblock {\em Cancers}, 13(17):4494, 2021.

\bibitem{wu2020classification}
Panpan Wu, Xuanchao Sun, Ziping Zhao, Haishuai Wang, Shirui Pan, and Bj{\"o}rn
  Schuller.
\newblock Classification of lung nodules based on deep residual networks and
  migration learning.
\newblock {\em Comput. Intell. Neurosci.}, 2020, 2020.

\bibitem{abbas2021acral}
Qaiser Abbas, Farheen Ramzan, and Muhammad~Usman Ghani.
\newblock Acral melanoma detection using dermoscopic images and convolutional
  neural networks.
\newblock {\em Vis. Comput. Ind. Biomed. Art}, 4(1):1--12, 2021.

\bibitem{bioengineering9030097}
Solene Bechelli and Jerome Delhommelle.
\newblock Machine learning and deep learning algorithms for skin cancer
  classification from dermoscopic images.
\newblock {\em Bioengineering}, 9(3), 2022.

\bibitem{alqudah2021sliding}
Amin Alqudah and Ali~Mohammad Alqudah.
\newblock Sliding window based deep ensemble system for breast cancer
  classification.
\newblock {\em J. Med. Eng. Technol.}, 45(4):313--323, 2021.

\bibitem{song2020end}
Lei Song, Jianzhe Lin, Z~Jane Wang, and Haoqian Wang.
\newblock An end-to-end multi-task deep learning framework for skin lesion
  analysis.
\newblock {\em IEEE J. Biomed. Health Inform.}, 24(10):2912--2921, 2020.

\bibitem{serte2020wavelet}
Sertan Serte and Hasan Demirel.
\newblock Wavelet-based deep learning for skin lesion classification.
\newblock {\em IET Image Process.}, 14(4):720--726, 2020.

\bibitem{foahom2021ensemble}
Arthur~Cartel Foahom~Gouabou, Jean-Luc Damoiseaux, Jilliana Monnier, Rabah
  Iguernaissi, Abdellatif Moudafi, and Djamal Merad.
\newblock Ensemble method of convolutional neural networks with directed
  acyclic graph using dermoscopic images: Melanoma detection application.
\newblock {\em Sensors}, 21(12):3999, 2021.

\bibitem{nasiri2020depict}
Sara Nasiri, Julien Helsper, Matthias Jung, and Madjid Fathi.
\newblock Depict melanoma deep-class: a deep convolutional neural networks
  approach to classify skin lesion images.
\newblock {\em BMC Bioinform.}, 21(2):1--13, 2020.

\bibitem{aydin2021detection}
Nevin Ayd{\i}n, {\"O}zer {\c{C}}elik, Ahmet~F Aslan, Alper Odaba{\c{s}}, Emine
  D{\"u}ndar, and Meryem~Cansu {\c{S}}ahin.
\newblock Detection of lung cancer on computed tomography using artificial
  intelligence applications developed by deep learning methods and the
  contribution of deep learning to the classification of lung carcinoma.
\newblock {\em Curr. Med. Imaging}, 17(9):1137--1141, 2021.

\bibitem{tan2019expert}
Jiaxing Tan, Yumei Huo, Zhengrong Liang, and Lihong Li.
\newblock Expert knowledge-infused deep learning for automatic lung nodule
  detection.
\newblock {\em J. X-ray Sci. Technol.}, 27(1):17--35, 2019.

\bibitem{bonavita2020integration}
Ilaria Bonavita, Xavier Rafael-Palou, Mario Ceresa, Gemma Piella, Vicent Ribas,
  and Miguel A~Gonz{\'a}lez Ballester.
\newblock Integration of convolutional neural networks for pulmonary nodule
  malignancy assessment in a lung cancer classification pipeline.
\newblock {\em Comput. Meth. Prog. Bio.}, 185:105172, 2020.

\bibitem{ohno2020differentiation}
Yoshiharu Ohno, Kota Aoyagi, Atsushi Yaguchi, Shinichiro Seki, Yoshiko Ueno,
  Yuji Kishida, Daisuke Takenaka, and Takeshi Yoshikawa.
\newblock Differentiation of benign from malignant pulmonary nodules by using a
  convolutional neural network to determine volume change at chest ct.
\newblock {\em Radiology}, 296(2):432--443, 2020.

\bibitem{zhang2019toward}
Chao Zhang, Xing Sun, Kang Dang, Ke~Li, Xiao-wei Guo, Jia Chang, Zong-qiao Yu,
  Fei-yue Huang, Yun-sheng Wu, Zhu Liang, et~al.
\newblock Toward an expert level of lung cancer detection and classification
  using a deep convolutional neural network.
\newblock {\em Oncologist}, 24(9):1159--1165, 2019.

\bibitem{choudhury2021detecting}
Avishek Choudhury and Sunanda Perumalla.
\newblock Detecting breast cancer using artificial intelligence: Convolutional
  neural network.
\newblock {\em Technol. Health Care}, 29(1):33--43, 2021.

\bibitem{wahab2019transfer}
Noorul Wahab, Asifullah Khan, and Yeon~Soo Lee.
\newblock Transfer learning based deep cnn for segmentation and detection of
  mitoses in breast cancer histopathological images.
\newblock {\em Microscopy}, 68(3):216--233, 2019.

\bibitem{sebai2020maskmitosis}
Meriem Sebai, Xinggang Wang, and Tianjiang Wang.
\newblock Maskmitosis: a deep learning framework for fully supervised, weakly
  supervised, and unsupervised mitosis detection in histopathology images.
\newblock {\em Med. Biol. Eng. Comput.}, 58(7):1603--1623, 2020.

\bibitem{zhou2019weakly}
Juan Zhou, Lu-Yang Luo, Qi~Dou, Hao Chen, Cheng Chen, Gong-Jie Li, Ze-Fei
  Jiang, and Pheng-Ann Heng.
\newblock Weakly supervised 3d deep learning for breast cancer classification
  and localization of the lesions in mr images.
\newblock {\em J. Magn. Reson. Imaging}, 50(4):1144--1151, 2019.

\bibitem{alom2019breast}
Md~Zahangir Alom, Chris Yakopcic, Mst Nasrin, Tarek~M Taha, Vijayan~K Asari,
  et~al.
\newblock Breast cancer classification from histopathological images with
  inception recurrent residual convolutional neural network.
\newblock {\em J. digit. imaging}, 32(4):605--617, 2019.

\bibitem{ciritsis2019automatic}
Alexander Ciritsis, Cristina Rossi, Matthias Eberhard, Magda Marcon, Anton~S
  Becker, and Andreas Boss.
\newblock Automatic classification of ultrasound breast lesions using a deep
  convolutional neural network mimicking human decision-making.
\newblock {\em Eur. Radiol.}, 29(10):5458--5468, 2019.

\bibitem{borkowski2020fully}
Karol Borkowski, Cristina Rossi, Alexander Ciritsis, Magda Marcon, Patryk
  Hejduk, Sonja Stieb, Andreas Boss, and Nicole Berger.
\newblock Fully automatic classification of breast mri background parenchymal
  enhancement using a transfer learning approach.
\newblock {\em Medicine}, 99(29), 2020.

\bibitem{noorbakhsh2020deep}
Javad Noorbakhsh, Saman Farahmand, Sandeep Namburi, Dennis Caruana, David Rimm,
  Mohammad Soltanieh-ha, Kourosh Zarringhalam, Jeffrey~H Chuang, et~al.
\newblock Deep learning-based cross-classifications reveal conserved spatial
  behaviors within tumor histological images.
\newblock {\em Nat. Commun.}, 11(1):1--14, 2020.

\bibitem{park2020deep}
Sohee Park, Sang~Min Lee, Kyung~Hee Lee, Kyu-Hwan Jung, Woong Bae, Jooae Choe,
  and Joon~Beom Seo.
\newblock Deep learning-based detection system for multiclass lesions on chest
  radiographs: comparison with observer readings.
\newblock {\em Eur. Radiol.}, 30(3):1359--1368, 2020.

\bibitem{masud2021machine}
Mehedi Masud, Niloy Sikder, Abdullah-Al Nahid, Anupam~Kumar Bairagi, and
  Mohammed~A AlZain.
\newblock A machine learning approach to diagnosing lung and colon cancer using
  a deep learning-based classification framework.
\newblock {\em Sensors}, 21(3):748, 2021.

\bibitem{nam2019development}
Ju~Gang Nam, Sunggyun Park, Eui~Jin Hwang, Jong~Hyuk Lee, Kwang-Nam Jin,
  Kun~Young Lim, Thienkai~Huy Vu, Jae~Ho Sohn, Sangheum Hwang, Jin~Mo Goo,
  et~al.
\newblock Development and validation of deep learning--based automatic
  detection algorithm for malignant pulmonary nodules on chest radiographs.
\newblock {\em Radiology}, 290(1):218--228, 2019.

\bibitem{sies2020past}
Katharina Sies, Julia~K Winkler, Christine Fink, Felicitas Bardehle, Ferdinand
  Toberer, Timo Buhl, Alexander Enk, Andreas Blum, Albert Rosenberger, and
  Holger~A Haenssle.
\newblock Past and present of computer-assisted dermoscopic diagnosis:
  performance of a conventional image analyser versus a convolutional neural
  network in a prospective data set of 1,981 skin lesions.
\newblock {\em Eur. J. Cancer}, 135:39--46, 2020.

\bibitem{zanjani2019impact}
Farhad~Ghazvinian Zanjani, Svitlana Zinger, Bastian Piepers, Saeed Mahmoudpour,
  Peter Schelkens, et~al.
\newblock Impact of jpeg 2000 compression on deep convolutional neural networks
  for metastatic cancer detection in histopathological images.
\newblock {\em J. Med. Imaging}, 6(2):027501, 2019.

\bibitem{lujan2021nanochest}
Juan~Eduardo Luj{\'a}n-Garc{\'\i}a, Yenny Villuendas-Rey, Itzam{\'a}
  L{\'o}pez-Y{\'a}{\~n}ez, Oscar Camacho-Nieto, and Cornelio
  Y{\'a}{\~n}ez-M{\'a}rquez.
\newblock Nanochest-net: A simple convolutional network for radiological
  studies classification.
\newblock {\em Diagnostics}, 11(5):775, 2021.

\bibitem{masquelin2021wavelet}
Axel~H Masquelin, Nicholas Cheney, C~Matthew Kinsey, and Jason~HT Bates.
\newblock Wavelet decomposition facilitates training on small datasets for
  medical image classification by deep learning.
\newblock {\em Histochem. Cell Biol.}, 155(2):309--317, 2021.

\bibitem{aly2021yolo}
Ghada~Hamed Aly, Mohammed Marey, Safaa~Amin El-Sayed, and Mohamed~Fahmy Tolba.
\newblock Yolo based breast masses detection and classification in full-field
  digital mammograms.
\newblock {\em Comput. Meth. Prog. Bio.}, 200:105823, 2021.

\bibitem{al2020evaluation}
Mugahed~A Al-Antari, Seung-Moo Han, and Tae-Seong Kim.
\newblock Evaluation of deep learning detection and classification towards
  computer-aided diagnosis of breast lesions in digital x-ray mammograms.
\newblock {\em Comput. Meth. Prog. Bio.}, 196:105584, 2020.

\bibitem{agnes2020classification}
S~Akila Agnes, J~Anitha, S~Pandian, and J~Dinesh Peter.
\newblock Classification of mammogram images using multiscale all convolutional
  neural network (ma-cnn).
\newblock {\em J. Med. Sys.}, 44(1):1--9, 2020.

\bibitem{chauvie2020artificial}
St{\'e}phane Chauvie, Adriano De~Maggi, Ilaria Baralis, Federico Dalmasso,
  Paola Berchialla, Roberto Priotto, Paolo Violino, Federico Mazza, Giulio
  Melloni, and Maurizio Grosso.
\newblock Artificial intelligence and radiomics enhance the positive predictive
  value of digital chest tomosynthesis for lung cancer detection within sos
  clinical trial.
\newblock {\em Eur. Radiol.}, 30(7):4134--4140, 2020.

\bibitem{binczyk2021radiomics}
Franciszek Binczyk, Wojciech Prazuch, Pawe{\l} Bozek, and Joanna Polanska.
\newblock Radiomics and artificial intelligence in lung cancer screening.
\newblock {\em Transl. Lung Cancer Res.}, 10(2):1186, 2021.

\bibitem{wang2020classification}
Shudong Wang, Liyuan Dong, Xun Wang, and Xingguang Wang.
\newblock Classification of pathological types of lung cancer from ct images by
  deep residual neural networks with transfer learning strategy.
\newblock {\em Open Med.}, 15(1):190--197, 2020.

\bibitem{khan2019lungs}
Sajid~A Khan, Muhammad Nazir, Muhammad~A Khan, Tanzila Saba, Kashif Javed,
  Amjad Rehman, Tallha Akram, and Muhammad Awais.
\newblock Lungs nodule detection framework from computed tomography images
  using support vector machine.
\newblock {\em J. Med. Sys.}, 82(8):1256--1266, 2019.

\bibitem{krajnc2021breast}
Denis Krajnc, Laszlo Papp, Thomas~S Nakuz, Heinrich~F Magometschnigg, Marko
  Grahovac, Clemens~P Spielvogel, Boglarka Ecsedi, Zsuzsanna Bago-Horvath,
  Alexander Haug, Georgios Karanikas, et~al.
\newblock Breast tumor characterization using [18f] fdg-pet/ct imaging combined
  with data preprocessing and radiomics.
\newblock {\em Cancers}, 13(6):1249, 2021.

\bibitem{barsha2021automated}
Nusrat~Ameen Barsha, Aimon Rahman, and MRC Mahdy.
\newblock Automated detection and grading of invasive ductal carcinoma breast
  cancer using ensemble of deep learning models.
\newblock {\em Comput. Biol. Med.}, 139:104931, 2021.

\bibitem{krijgsman2020quantitative}
Dani{\"e}lle Krijgsman, Marinus~B van Leeuwen, John van~der Ven, Vanda Almeida,
  Ruud Vlutters, David Halter, Peter~JK Kuppen, Cornelis~JH van~de Velde, and
  Reinhold Wimberger-Friedl.
\newblock Quantitative whole slide assessment of tumor-infiltrating
  cd8-positive lymphocytes in er-positive breast cancer in relation to clinical
  outcome.
\newblock {\em IEEE J. Biomed. Health Inform.}, 25(2):381--392, 2020.

\bibitem{loizidou2021digital}
Kosmia Loizidou, Galateia Skouroumouni, Costas Pitris, and Christos Nikolaou.
\newblock Digital subtraction of temporally sequential mammograms for improved
  detection and classification of microcalcifications.
\newblock {\em Eur. Radiol. Exp.}, 5(1):1--12, 2021.

\bibitem{liang2021reducing}
Jinglun Liang, Guoliang Ye, Jianwen Guo, Shaohui Zhang, and QiFan Huang.
\newblock Reducing false-positives in lung nodules detection using balanced
  datasets.
\newblock {\em Front. Public Health}, 9:517, 2021.

\bibitem{saba2019automated}
Tanzila Saba.
\newblock Automated lung nodule detection and classification based on multiple
  classifiers voting.
\newblock {\em J. Med. Sys.}, 82(9):1601--1609, 2019.

\bibitem{santilli2021domain}
Alice~ML Santilli, Amoon Jamzad, Alireza Sedghi, Martin Kaufmann, Kathryn
  Logan, Julie Wallis, Kevin~YM Ren, Natasja Janssen, Shaila Merchant, Jay
  Engel, et~al.
\newblock Domain adaptation and self-supervised learning for surgical margin
  detection.
\newblock {\em Int. J. Comput. Ass. Rad.}, 16(5):861--869, 2021.

\bibitem{hosking2019hyperspectral}
Anna-Marie Hosking, Brandon~J Coakley, Dorothy Chang, Faezeh Talebi-Liasi,
  Samantha Lish, Sung~Won Lee, Amanda~M Zong, Ian Moore, James Browning,
  Steven~L Jacques, et~al.
\newblock Hyperspectral imaging in automated digital dermoscopy screening for
  melanoma.
\newblock {\em Lasers Surg. Med.}, 51(3):214--222, 2019.

\bibitem{wu2019machine}
Tong Wu, Laith~R Sultan, Jiawei Tian, Theodore~W Cary, and Chandra~M Sehgal.
\newblock Machine learning for diagnostic ultrasound of triple-negative breast
  cancer.
\newblock {\em Breast Cancer Res. Treat.}, 173(2):365--373, 2019.

\bibitem{marchetti2020computer}
Michael~A Marchetti, Konstantinos Liopyris, Stephen~W Dusza, Noel~CF Codella,
  David~A Gutman, Brian Helba, Aadi Kalloo, Allan~C Halpern, H~Peter Soyer,
  Clara Curiel-Lewandrowski, et~al.
\newblock Computer algorithms show potential for improving dermatologists'
  accuracy to diagnose cutaneous melanoma: Results of the international skin
  imaging collaboration 2017.
\newblock {\em J. Am. Acad. Dermatol.}, 82(3):622--627, 2020.

\bibitem{maron2020artificial}
Roman~C Maron, Jochen~S Utikal, Achim Hekler, Axel Hauschild, Elke Sattler,
  Wiebke Sondermann, Sebastian Haferkamp, Bastian Schilling, Markus~V Heppt,
  Philipp Jansen, et~al.
\newblock Artificial intelligence and its effect on dermatologists’ accuracy
  in dermoscopic melanoma image classification: web-based survey study.
\newblock {\em J. Med. Internet Res.}, 22(9):e18091, 2020.

\bibitem{wu2019deep}
Nan Wu, Jason Phang, Jungkyu Park, Yiqiu Shen, Zhe Huang, Masha Zorin,
  Stanis{\l}aw Jastrzebski, Thibault F{\'e}vry, Joe Katsnelson, Eric Kim,
  et~al.
\newblock Deep neural networks improve radiologists’ performance in breast
  cancer screening.
\newblock {\em IEEE Trans. Med. Imag.}, 39(4):1184--1194, 2019.

\bibitem{choi2021evaluation}
Soo~Yun Choi, Sunggyun Park, Minchul Kim, Jongchan Park, Ye~Ra Choi, and
  Kwang~Nam Jin.
\newblock Evaluation of a deep learning-based computer-aided detection
  algorithm on chest radiographs: Case--control study.
\newblock {\em Medicine}, 100(16), 2021.

\bibitem{wang2020inconsistent}
Xiaoqin Wang, Gongbo Liang, Yu~Zhang, Hunter Blanton, Zachary Bessinger, and
  Nathan Jacobs.
\newblock Inconsistent performance of deep learning models on mammogram
  classification.
\newblock {\em J. Am. Coll. Radiol.}, 17(6):796--803, 2020.

\bibitem{tennakoon2019classification}
Ruwan Tennakoon, Gerda Bortsova, Silas {\O}rting, Amirali~K Gostar, Mathilde~MW
  Wille, Zaigham Saghir, Reza Hoseinnezhad, Marleen de~Bruijne, and Alireza
  Bab-Hadiashar.
\newblock Classification of volumetric images using multi-instance learning and
  extreme value theorem.
\newblock {\em IEEE Trans. Med. Imag.}, 39(4):854--865, 2019.

\bibitem{ibrahim2021deep}
Dina~M Ibrahim, Nada~M Elshennawy, and Amany~M Sarhan.
\newblock Deep-chest: Multi-classification deep learning model for diagnosing
  covid-19, pneumonia, and lung cancer chest diseases.
\newblock {\em Comput. Biol. Med.}, 132:104348, 2021.

\bibitem{ektefaie2021integrative}
Yasha Ektefaie, William Yuan, Deborah~A Dillon, Nancy~U Lin, Jeffrey~A Golden,
  Isaac~S Kohane, and Kun-Hsing Yu.
\newblock Integrative multiomics-histopathology analysis for breast cancer
  classification.
\newblock {\em NPJ Breast Cancer}, 7(1):1--6, 2021.

\bibitem{kehl2019assessment}
Kenneth~L Kehl, Haitham Elmarakeby, Mizuki Nishino, Eliezer~M Van~Allen, Eva~M
  Lepisto, Michael~J Hassett, Bruce~E Johnson, and Deborah Schrag.
\newblock Assessment of deep natural language processing in ascertaining
  oncologic outcomes from radiology reports.
\newblock {\em JAMA Oncol.}, 5(10):1421--1429, 2019.

\bibitem{oei2019convolutional}
Ronald~Wihal Oei, Guanqun Hou, Fuhai Liu, Jin Zhong, Jiewen Zhang, Zhaoyi An,
  Luping Xu, and Yujiu Yang.
\newblock Convolutional neural network for cell classification using microscope
  images of intracellular actin networks.
\newblock {\em PLoS one}, 14(3):e0213626, 2019.

\bibitem{wang2021deep}
Chengdi Wang, Xiuyuan Xu, Jun Shao, Kai Zhou, Kefu Zhao, Yanqi He, Jingwei Li,
  Jixiang Guo, Zhang Yi, and Weimin Li.
\newblock Deep learning to predict egfr mutation and pd-l1 expression status in
  non-small-cell lung cancer on computed tomography images.
\newblock {\em J. Oncol.}, 2021, 2021.

\bibitem{wang2021prediction}
Xiaoxiao Wang, Chong Zou, Yi~Zhang, Xiuqing Li, Chenxi Wang, Fei Ke, Jie Chen,
  Wei Wang, Dian Wang, Xinyu Xu, et~al.
\newblock Prediction of brca gene mutation in breast cancer based on deep
  learning and histopathology images.
\newblock {\em Front. Genet.}, page 1147, 2021.

\bibitem{anand2020deep}
Deepak Anand, Nikhil~Cherian Kurian, Shubham Dhage, Neeraj Kumar, Swapnil Rane,
  Peter~H Gann, and Amit Sethi.
\newblock Deep learning to estimate human epidermal growth factor receptor 2
  status from hematoxylin and eosin-stained breast tissue images.
\newblock {\em J. Pathol. Inform.}, 11, 2020.

\bibitem{al2021nanogenomics}
Batla~S Al-Sowayan and Alaa~T Al-Shareeda.
\newblock Nanogenomics and artificial intelligence: A dynamic duo for the fight
  against breast cancer.
\newblock {\em Frontiers in Molecular Biosciences}, Front. Mol. Biosci.:219,
  2021.

\bibitem{han2020augmented}
Seung~Seog Han, Ilwoo Park, Sung~Eun Chang, Woohyung Lim, Myoung~Shin Kim,
  Gyeong~Hun Park, Je~Byeong Chae, Chang~Hun Huh, and Jung-Im Na.
\newblock Augmented intelligence dermatology: deep neural networks empower
  medical professionals in diagnosing skin cancer and predicting treatment
  options for 134 skin disorders.
\newblock {\em J. Invest. Dermatol.}, 140(9):1753--1761, 2020.

\bibitem{ho2021detecting}
Chi-Jui Ho, Manuel Calderon-Delgado, Chin-Cheng Chan, Ming-Yi Lin, Jeng-Wei
  Tjiu, Sheng-Lung Huang, and Homer~H Chen.
\newblock Detecting mouse squamous cell carcinoma from submicron full-field
  optical coherence tomography images by deep learning.
\newblock {\em J. Biophotonics}, 14(1):e202000271, 2021.

\bibitem{kumar2021integration}
Rajesh Kumar, WenYong Wang, Jay Kumar, Ting Yang, Abdullah Khan, Wazir Ali, and
  Ikram Ali.
\newblock An integration of blockchain and ai for secure data sharing and
  detection of ct images for the hospitals.
\newblock {\em Comput. Med. Imag. Graph.}, 87:101812, 2021.

\bibitem{udrea2020accuracy}
Andreaa Udrea, GD~Mitra, D~Costea, EC~Noels, Marlies Wakkee, DM~Siegel,
  TM~de~Carvalho, and TEC Nijsten.
\newblock Accuracy of a smartphone application for triage of skin lesions based
  on machine learning algorithms.
\newblock {\em J. Eur. Acad. Dermatol. Venereol.}, 34(3):648--655, 2020.

\bibitem{jiang2020recognizing}
YQ~Jiang, JH~Xiong, HY~Li, XH~Yang, WT~Yu, M~Gao, X~Zhao, YP~Ma, W~Zhang,
  YF~Guan, et~al.
\newblock Recognizing basal cell carcinoma on smartphone-captured digital
  histopathology images with a deep neural network.
\newblock {\em Br. J. Dermatol.}, 182(3):754--762, 2020.

\end{thebibliography}

\end{document}